\documentclass[journal]{vgtc}                     % final (journal style)
\onlineid{1096}

\vgtccategory{Research}

\vgtcpapertype{applicaition}

\title{Fine-Tuned Large Language Model for Visualization System: A Study on Self-Regulated Learning in Education}

\author{%
  \authororcid{Lin Gao}{0009-0004-1613-1774},
  Jing Lu, 
  \authororcid{Zekai Shao}{0000-0003-2014-5293}, 
  Ziyue Lin, 
  \authororcid{Shengbin Yue}{0000-0002-6764-1756}, 
  Chiokit Ieong, 
  Yi Sun, \\
  Rory James Zauner, 
  \authororcid{Zhongyu Wei}{0000-0003-3789-8507} and 
  \authororcid{Siming Chen}{0000-0002-2690-3588}
}

\authorfooter{
    \item
  	Lin Gao, Jing Lu, Zekai Shao, Ziyue Lin, Shengbin Yue, Chiokit Ieong, Yi Sun, Zhongyu Wei, and Siming Chen are with the School of Data Science at Fudan University. S. Chen is also with the Shanghai Key Laboratory of Data Science. S. Chen is the corresponding author.
  	E-mail: lingao23@m.fudan.edu.cn, simingchen@fudan.edu.cn
   \item 
    Rory James Zauner is with the Faculty of Computer Science, University of Vienna, Vienna, Austria.
}

\abstract{%
Large Language Models (LLMs) have shown great potential in intelligent visualization systems, especially for domain-specific applications. Integrating LLMs into visualization systems presents challenges, and we categorize these challenges into three alignments: domain problems with LLMs, visualization with LLMs, and interaction with LLMs. To achieve these alignments, we propose a framework and outline a workflow to guide the application of fine-tuned LLMs to enhance visual interactions for domain-specific tasks. These alignment challenges are critical in education \textcolor{revision}{because of the need for} an intelligent visualization system to support beginners' self-regulated learning. Therefore, we apply the framework to education and introduce Tailor-Mind, an interactive visualization system designed to facilitate self-regulated learning for artificial intelligence beginners. Drawing on insights from a preliminary study, we identify self-regulated learning tasks and fine-tuning objectives to guide visualization design and tuning data construction. Our focus on aligning visualization with fine-tuned LLM makes Tailor-Mind more like a personalized tutor. Tailor-Mind also supports interactive recommendations to help beginners better achieve their learning goals. Model performance evaluations and user studies confirm that Tailor-Mind improves the self-regulated learning experience, effectively validating the proposed framework.

}


\teaser{
  \centering
  \includegraphics[width=\linewidth]{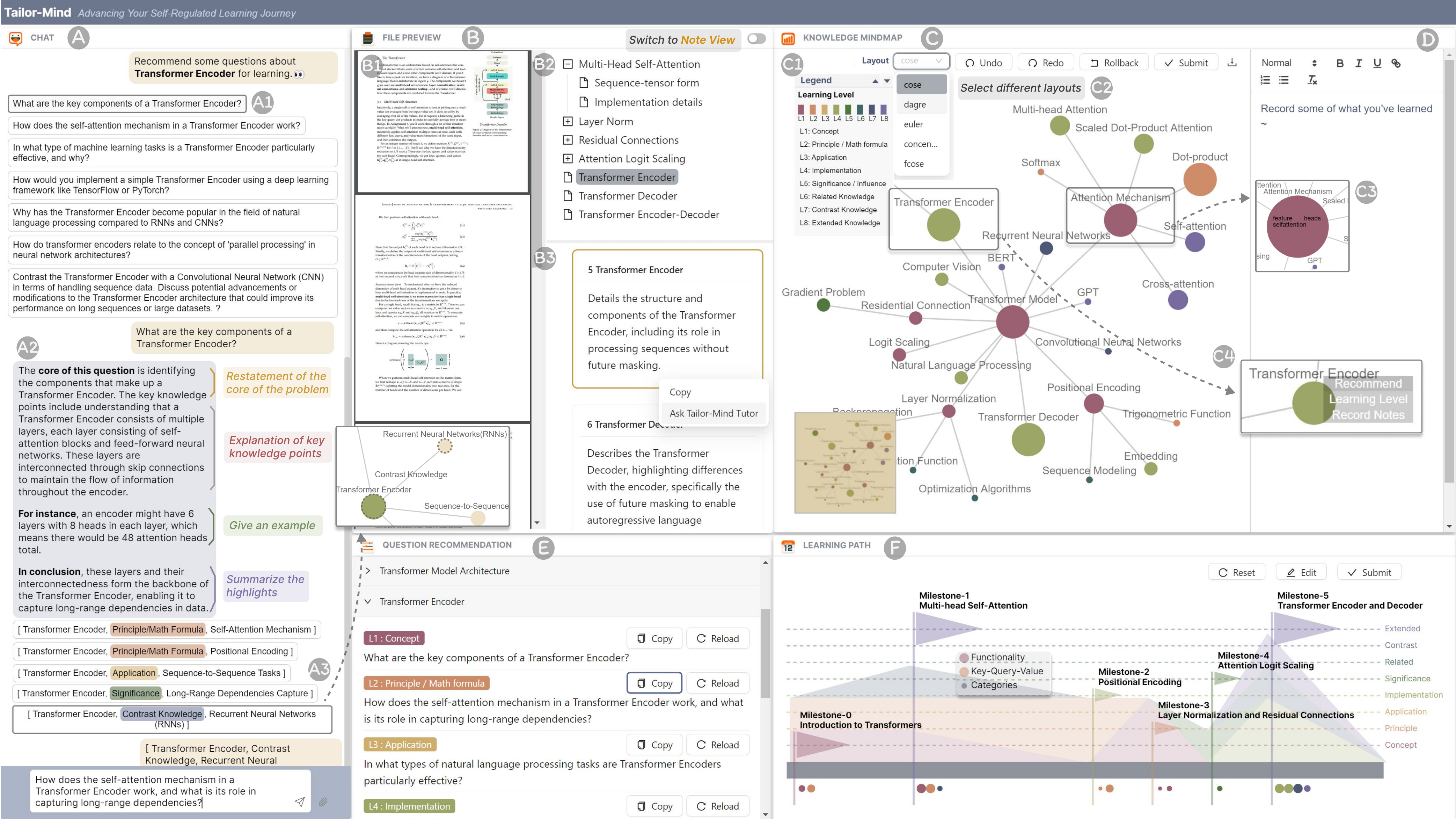}
  \caption{%
  The overview of \textit{Tailor-Mind.} The Chat View (A) enables users to engage in conversations and receive recommended questions (A1), enhancing their knowledge exploration. Intelligent responses provide in-depth explanations (A2) and guide users to discover new knowledge connections (A3). The File Preview (B) displays learning materials (B1), knowledge structures in a tree widget (B2), and knowledge cards (B3). The Knowledge MindMap (C) visualizes the network of knowledge based on the learning levels (C1), allowing users to select different layouts (C2). Knowledge nodes display users' notes (C3), support recommended questions, and allow for the modification of learning goals (C4). Users are also encouraged to take notes in a rich text editor (D). The Question Recommendation (E) offers questions about extracted knowledge points on various learning levels. The Learning Path (F) helps users understand their learning plans and delve into each detailed learning objective. 
  }
  \label{fig:teaser}
}

\usepackage{tabu}                      % only used for the table example
\usepackage{booktabs}                  % only used for the table example
\usepackage{lipsum}                    % used to generate placeholder text
\usepackage{mwe}                       % used to generate placeholder figures
\usepackage{mathptmx}                  % use matching math font
\usepackage{color}
\usepackage{enumitem}
\usepackage{graphicx}
\usepackage{multirow}                  % used to show table
\usepackage{tabularx}
\usepackage{setspace}
\usepackage[normalem]{ulem}
\usepackage{wrapfig}
\usepackage{url}
\definecolor{remark}{RGB}{197, 90, 17}
\definecolor{revision}{RGB}{0,0,0}

\begin{document}

\maketitle
%%%%%%%%%%%%%%%%%%%%%%%%%%%%%%%%%%%%%%%%%%%%%%%%%%%%%%%%%%%%%%%%
%%%%%%%%%%%%%%%%%%%%%% START OF THE PAPER %%%%%%%%%%%%%%%%%%%%%%
%%%%%%%%%%%%%%%%%%%%%%%%%%%%%%%%%%%%%%%%%%%%%%%%%%%%%%%%%%%%%%%%


\begin{spacing}{0.97}
\section{Introduction} \label{sec:introduction}
%% Background
% LLM supports domain-specific complex tasks
The rapid advancement of large language models (LLMs) has captured significant attention~\cite{openai2024gpt4,zeng2023glmb,touvron2023llama} and opened new avenues for tackling specialized domain problems~\cite{ling2023domain,liu2022semiconductors}. Increasingly, researchers are applying these general models to specific areas, with supervised fine-tuning~\cite{wei2022finetuned} emerging as a practical approach. By training on domain-specific datasets, this method enhances the model's performance in fields such as healthcare~\cite{thirunavukarasu2023large}, legal~\cite{yue2023disclawllm}, and education~\cite{dan2023educhat}. Domain-specific LLMs, fine-tuned in this way, demonstrate improved decision-making, knowledge integration, retrieval, and logical reasoning, addressing issues of instability and inaccuracy commonly found in general models when applied to domain-specific tasks~\cite{zhang2022fine-tuning}.
% Visualization supports domain-specific complex tasks 
Visualization systems are also commonly used to solve domain-specific problems because they reveal data insights and provide a good user experience through interactions. Some researchers have specifically proposed general frameworks in the visualization community to guide analysts in designing and implementing visualization workflows for domain tasks~\cite{Keim2008visual,andrienko2018viewing}. However,  traditional visualization systems that incorporate small-scale models have limited capabilities for intelligent analysis. Due to the intelligence of LLMs, many researchers are considering integrating LLMs into the visual interaction processes of domain-specific problems.

% Transition to present the problems 
However, integrating LLMs and visualization systems to address domain-specific problems encounters some difficulties, owing to their distinct capabilities, which complicate the optimal leveraging of their respective strengths.
%% alignment problems
% 1. Alignment between domain problems and large language models. => adaptation of LLMs to domain-specific problems
\textcolor{revision}{The first challenge is the adaptation of LLMs to domain-specific problems. For the strong domain barriers in these problems, general LLMs mostly lack the ability to handle specific, complex domain knowledge and therefore cannot effectively solve domain problems.}
% 2. Alignment between visualizations and large language models.=> synchronization between visualization and LLMs
\textcolor{revision}{The second challenge is synchronizing visualization with LLMs. LLMs lack knowledge of visualization and an understanding of visualization systems. To empower visualization systems with LLM capabilities, it is crucial to ensure that models internalize the visualization process and visual mappings as knowledge and actions.}
% 3. Alignment between interactions and large language models.=> understanding and enhancing interactions with LLMs
\textcolor{revision}{The third challenge involves understanding and enhancing interactions with LLMs. LLMs need to understand various types of interactions and how to provide personalized interactive recommendations. Ensuring LLMs accurately interpret user intentions from different interactions is vital for achieving effective and personalized interactive experiences.}

% Current research
% 现有的LLM-empowered visualization system的工作可以利用LLM的能力智能化、自动化的解决很多问题。但是他们只是将LLM的能力单纯的嵌入到系统中，对于复杂的特定领域知识缺乏模型知识和行为的改进。
\textcolor{revision}{Existing LLM-empowered visualization systems can intelligently and automatically solve problems~\cite{hou2024c2ideas, wang2024virtuwander}. However, they merely embed LLM into the system without enhancing the model's knowledge and behavior for complex, domain-specific knowledge.} 
% 同时，也有一些工作通过微调在解决领域问题，但是整个构建的过程与可视化中支持的任务以及整个用户可视化交互流程是割裂的。
\textcolor{revision}{Additionally, some efforts address domain-specific problems by fine-tuning~\cite{wang2023commonsensevis,huang2024plantography}. However, these fine-tuned models are constructed in isolation from the visualization process and do not consider the tasks and requirements involved in visualization.}
% 目前，缺少相关工作综合考虑领域知识、可视化与交互在大语言模型智能解决特定领域问题上的结合。我们需要进一步分析它们之间的关系，识别一些模式，以指导更好地解决领域任务。
% Therefore, integrating LLMs into visualization systems for domain applications goes beyond simply combining the model with visualization systems.
\textcolor{revision}{Therefore, there is a lack of work that integrates domain knowledge, visualization, and interaction to leverage LLMs for domain-specific problems.}
% We need to analyze their relationships and identify patterns to guide better solutions.}
% Current LLM-empowered visualization systems have not yet fully covered these three alignments~\cite{hou2024c2ideas, wang2024virtuwander}. Efforts should contend with the instability and inaccuracies in LLM outputs~\cite{wu2022promptchain}. 
% There exists a gap between LLMs' grasp of domain knowledge and their ability to represent it visually. 
% % Significance & transition
% Therefore, integrating LLMs into visualization systems for domain applications goes beyond simply combining the model with visualization systems.

%% Contributions
% Framework
% 在本文中，我们分析LLMs和这些元素之间的关系与模式，总结出三个对齐目标。并基于这三个对齐目标，我们提出了一个概念性框架来指导针对可视化要解决的领域任务能够做tuning，这样的能力通过交互探索背后大模型的支持完成领域任务。
In this paper, \textcolor{revision}{we analyze the relationships and patterns among LLMs, domain knowledge, visualization, and interaction, identifying three alignment objectives: domain problems with LLMs, visualizations with LLMs, and interactions with LLMs. Based on these alignments, we propose a conceptual framework to guide the tuning of LLMs for domain-specific tasks in visualization systems.}
% \textcolor{revision}{we analyze the relationships among domain tasks and data, visualization systems, fine-tuned LLMs, and user interactions under the three alignments. We propose a framework that guides the general workflow with detailed guidelines.} 
\textcolor{revision}{We apply our proposed framework to the educational domain, introducing Tailor-Mind for artificial intelligence (AI) beginners. A preliminary study with students and teachers reveals a need for intelligent interactive systems in Self-Regulated Learning (SRL), which our approach aims to address.} \textcolor{revision}{By integrating expert insights with user needs, learning tasks are defined, leading to the establishment of design requirements and fine-tuning objectives for the domain model. This optimization ensures the model fits the learning tasks and maintains interactive functionality between visualization and students' learning process.} To our knowledge, this work presents the first conceptual framework that combines fine-tuned models and visualization systems to tackle domain-specific problems. 
The contributions are as follows:
\begin{itemize}[parsep=0.08pt]
  \item \textbf{A conceptual framework} that integrates fine-tuned LLMs into interactive visualization systems, alongside \textbf{a workflow} of applying the framework to different domains.
  
  \item Applying the framework to the education domain, we introduce \textbf{Tailor-Mind}, an interactive visualization system for artificial intelligence beginners supported by a fine-tuned LLM. The system supports intelligent exploration of knowledge and personalized recommendations during the self-regulated learning process.
  
  \item The evaluation of model performance, alongside findings from usage scenarios and user study, validates Tailor-Mind’s effectiveness in facilitating SRL experiences. This substantiates the framework’s rationality and feasibility and offers the educational domain valuable insights for promoting active, iterative learning.
\end{itemize}
\section{Related Work}\label{sec:related_work}
This section reviews related work on fine-tuned LLMs for domain-specific applications and LLM-empowered visualization systems, especially in the education domain.

\subsection{LLM-based Visualization System for Domain Tasks}

\textcolor{revision}{The exceptional performance of LLMs has facilitated their integration into visual interactions. Currently, their power can be leveraged in several ways: external knowledge bases, prompt engineering, agent settings, and data fine-tuning.}

\textcolor{revision}{\textbf{External knowledge bases.} External knowledge bases enhance LLMs by providing vast repositories of structured information~\cite{lewis2020rag}.} DocFlow~\cite{qiu2023docflow} intelligently classifies documents by incorporating document information retrieval techniques based on user questions. Peng et al.~\cite{peng2023check} proposed an LLM-AUGMENTER system that augments LLM with plug-and-play modules based on external knowledge. 

\textcolor{revision}{\textbf{Prompt engineering.}} Through prompt engineering, LEVA~\cite{Yuheng2024LEVA} enables LLMs to generate the declarative syntax of a visual analytics system, understand view relationships, and interpret diagram information to provide analysis tasks and interaction recommendations. \textcolor{revision}{Works tailored to specific scenarios, such as interior design~\cite{hou2024c2ideas} and virtual museum tours~\cite{wang2024virtuwander}, enhance user experiences by interpreting inputs and adapting outputs, thereby facilitating intelligent and personalized solutions. Simultaneously, there is work~\cite{strobelt2022interactive} that focuses on visualizing prompt performance and methods for iterative optimization.} 
% C2Ideas~\cite{hou2024c2ideas} hones user intent, interprets meanings, and adapts design principles to offer bespoke interior design solutions. virtuWander~\cite{wang2024virtuwander} converts user inquiries into varied contexts to provide guidance, enabling multi-modal interactions for tailored virtual museum tours. 

\textcolor{revision}{\textbf{Agent setting.} A unique application of prompt projects lies in agent design, where the LLM-based conversational agent enhances user immersion in conjunction with interaction design~\cite{wu2024socrates, liu2024how}. 
The sandbox is another scenario where LLM-based agents are used to simulate social behaviors~\cite{park2023generative}. AgentLens~\cite{lu2024agentlens} illustrates the evolution of LLM-based autonomous systems through hierarchical temporal visualization.} 

\textcolor{revision}{\textbf{Data fine-tuning.} Several systems employ data fine-tuning to personalize models, primarily focusing on addressing natural language questions.} 
CommonsenseVIS~\cite{wang2023commonsensevis} adds concept and relation alignments to improve model behavior contextualization and question-answering ability.
PlantoGraphy~\cite{huang2024plantography} uses a fine-tuned model to transform garden scene layouts into realistic landscape renderings. 
% Several systems employ data fine-tuning to personalize models and primarily focus on referring natural language questions.

The above work demonstrates that integrating LLMs into visualization systems improves user experience. LLM performance in domain tasks is becoming increasingly critical, and the challenge of enhancing domain task completion through the LLM-empowered visualization system remains a concern. Our work focuses on aligning model performance with visualization and interaction to assist domain task-solving.

\subsection{Fine-tuning LLMs for Domain Specific Applications}

The advent of LLMs like ChatGPT~\cite{openai2023chatgpt} and LLaMa~\cite{touvron2023llama} has catalyzed research into leveraging their formidable powers across diverse professional domains. Wei et al.~\cite{wei2022finetuned} introduced an innovative instruction fine-tuning technique to bolster the models' adaptability to domain tasks~\cite{zhang2023instruction}. In the judiciary, projects like DISC-LawLLM~\cite{yue2023disclawllm} has made significant headway in addressing intricate tasks such as legal element identification, case sorting, and judgment forecasting. In the financial sector, FinGPT~\cite{yang2023fingpt} stands as a testament to the potential of large models developed through a thorough analysis of financial narratives, social discourse, and fiscal reports. In healthcare, BenTsao~\cite{wang2023huatuo} has improved models' question-answering capabilities and proposed a fine-tuned dataset~\cite{du2023calla} leveraging technologies like knowledge graphs~\cite{wang2023knowledgetuning}. Initiatives such as EduChat~\cite{dan2023educhat} and Taoli Llama~\cite{jingsi2023taoli} have enhanced the utility of large models, meeting the growing call for accessible models in the educational sphere~\cite{tack2023bea,baladon2023retuyt}. Moreover, in specific disciplines, Yue et al.~\cite{yue2023mammoth} trained the MAmmoTH series of models with enhanced mathematical reasoning ability by mixing Chain of Thought (CoT)~\cite{Wei2022ChainOT} and Programming of Thought (PoT)~\cite{chen2023program}. PromptProtein~\cite{wang2023multilevel} focused on protein sequence prediction, demonstrating the value of discipline-specific LLMs.

Instruction fine-tuning is acknowledged for enhancing model efficacy and adaptability in targeted domains. Yet, developing fine-tuning datasets is intricate and laborious, with the dataset's quality and volume being pivotal to the model's domain-specific outcomes~\cite{wang2023selfinstruct, schick2021generating}. For many developers, constructing datasets for fine-tuning presents a formidable and time-intensive challenge~\cite{li2023quantity, schick2023toolformer}.

\subsection{Intelligent Tutorial System and Educational Agent}

Since the illustrating workflow pertains to education, it is essential to analyze how previous studies utilized LLM alongside visualization to augment the pedagogical landscape~\cite{hwang2020vision}. A crucial aspect of an intelligent tutorial system (ITS) is interactive visualization, offering an intuitive and engaging educational journey~\cite{crow2018intelligent}. Wang et al.~\cite{zheng2022telling} have innovated by automating slide generation from Jupyter Notebooks. TransforLearn~\cite{gao2023transforlearn} provides an interactive way to understand the Transformer model and data flows. Kabdo et al.~\cite{choi2022algosolve} have proposed AlgoSolve, a tool aiding learners in algorithmic problems. The advancement of LLMs has given rise to LLM-empowered ITS~\cite{kasneci2023chatgpt}. Storyfier\cite{peng2023storyfier} utilizes LLMs in language practice, generating contexts that encompass target words. UKP-SQuARE~\cite{fang2023ukp} offers a platform to operate, assess, and analyze various QA models. Another critical area in ITS is using educational agents to instruct, motivate, and engage learners~\cite{xi2023rise}. HypoCompass~\cite{ma2023hypocompass} proposes a learning-by-teaching approach, where one agent functions as a student, while a novice oversees the debugging process. Ruffle \& Riley~\cite{schmucker2023ruffle} demonstrate student and teacher roles, executing tutorial scripts from textbooks.

LLM-enhanced approaches offer novel learning solutions yet often fail to meet domain-specific standards, with effectiveness tied to the model's capabilities. Moreover, these methods primarily act as chatbots, lacking comprehensive learner guidance. In educational theory, fostering students' SRL capabilities is paramount~\cite{panadero2017review}. Consequently, we integrate the proposed framework within the educational context, constructing instructional datasets for fine-tuned domain models and merging the model with a visualization system to assist users in accomplishing SRL tasks.
\section{Fine-Tuned LLM for Visualization System} \label{sec: framework&workflow}
In Sec.~\ref{sec: problem definition}, the challenges of integrating LLMs into visualization systems are summarized in three alignments. To address these challenges, we discuss the proposed framework (Sec.~\ref{sec: framework}) and analyze how it can be applied to solve domain problems (Sec.~\ref{sec: workflow}).

\subsection{Problem Definition} \label{sec: problem definition}
%% background
% Due to the user experience, visual interaction is the preferred way to deal with problems~\cite{liu2021supporting}. With the high intelligence of LLMs in the spotlight, improving the intelligence of visualization systems is also an important issue. However, integrating LLMs into visualization systems presents some challenges. We summarize the challenge as three alignments:
% Due to user preferences for visual interaction~\cite{liu2021supporting}, enhancing the intelligence of visualization systems is crucial as LLMs capabilities expand. 
Given user preferences for visual interaction~\cite{liu2021supporting}, enhancing the intelligence of visualization systems becomes crucial.
However, integrating LLMs into visualization systems presents multiple challenges. To tackle these, we propose three alignments. 
\textcolor{revision}{We define ``alignment'' as the mutual adaptation and coordination between LLMs and various elements such as visualization systems, domain knowledge, and user interactions, aimed at meeting multifaceted requirements and optimizing performance for models' targeted behaviors.}

\begin{enumerate}[label=A\arabic*,itemsep=0pt, parsep=0pt]
\item \label{A1} \textbf{Alignment between domain problems and LLMs.} 
It is critical to integrate domain-specific knowledge (e.g., terminology and concepts), experiences, and insights into the LLM and match the linguistic patterns. This helps LLMs to skillfully solve complex tasks~\cite{he2024large} that require a deep understanding of domain specificity.

\item \label{A2} \textbf{Alignment between visualizations and LLMs.} It's essential to account for the context in which visualization operates, including the specific data features or insights. The visualization design and its alignment with model outputs are paramount. This design extends beyond visual representation~\cite{shi2023reverse} to encompass the system's pipeline to reflect domain-specific problem-solving. 

\item \label{A3} \textbf{Alignment between interactions and LLMs.} LLMs should interpret user intents behind interactions, including natural language, non-verbal cues, and visual elements. Furthermore, since user interactions can be aimless and uncertain, LLMs need to adapt to exploratory queries and offer guided hints.
\end{enumerate}

\subsection{Framework}\label{sec: framework}
\begin{figure}[h]
\vspace{-0.3cm}
  \centering 
  \includegraphics[width=\columnwidth]{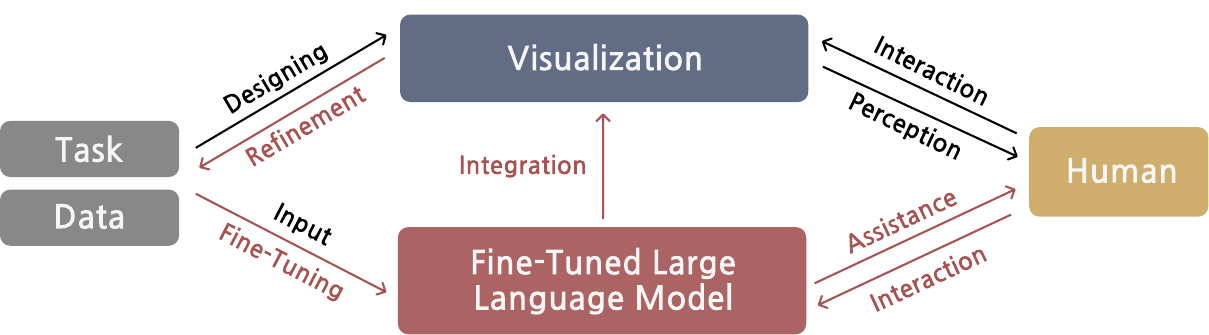}
  \caption{Framework of integrating fine-tuned LLM into visualization system. The black lines indicate relationships in traditional processes, and the red lines highlight connections that are introduced or altered by the involvement of fine-tuned LLM.}
  \label{fig: framework}
  \vspace{-0.3cm}
\end{figure}

Based on the traditional framework of visual analytics processes~\cite{Keim2008visual, andrienko2018viewing}, we think of the impact of the LLM's addition and propose the conceptual framework, as illustrated in Fig.~\ref{fig: framework}. In this framework, we use domain tasks and data as inputs to provide people with task solution results and experiences through an intelligent visualization system based on the fine-tuned LLM.

\textbf{Node Description.} \textit{Task} possesses strong domain-specific traits and is typically a complex domain problem consisting of multiple sub-tasks. \textit{Data} consists of the data provided by users and the data needed to be transformed into domain knowledge. \textit{Human} is the subject involved in the whole process. Additionally, humans are the end users of the fine-tuned LLM and visualization system, and they interact with them to accomplish the task. \textit{Visualization} refers to the visualization system, which provides interaction and visual perception. Humans use the visualization system as a platform to complete tasks. \textit{Fine-tuned LLM} outperforms the small-scale model and excels at handling domain tasks than the generalized LLM.

%% alignment 1
\textbf{\textit{Fine-tuning} aligns LLM with domain requirements.} 
Fine-tuning requires that models learn to ``know'' domain knowledge and ``apply'' knowledge to solve problems. 
% know domain knowledge
Converting domain data into high-quality knowledge is a key aspect of fine-tuning. High-quality data for fine-tuning can help models distill insights and knowledge from it. 
% apply knowledge to solve problems
Moreover, fine-tuning involves teaching LLM the processing flow and expertise from domain-specific tasks. The model needs to develop behaviors for handling domain-specific problems by comprehending the logic and sequence between sub-tasks rather than processing a task in isolation. 

%% alignment 2
\textbf{\textit{Integration} and \textit{refinement} of LLMs with visualization systems.}
% integration process
In the process of integration, the fine-tuned model needs to understand the original intent of the visualization design, including the design purposes, data features, and communications of relevant insights. The model also needs to learn the contextual logic required for the insights to align the visualization. 
% refinement process
Refinement strategies can be extracted from visual designs and system workflows, and they can be applied to facilitate model handling of domain problems in the visualization system.

%% alignment 3
\textbf{Intelligent \textit{assistance} and \textit{interaction} from LLM align with user requirements.}
% intelligent assistance
The fine-tuned LLM aligns with human preferences, habits, and behavioral patterns. This adaptation enables the LLM to seamlessly deliver intelligent assistance during user interactions, whether through natural language processing or visual interface engagement.
% guided interaction
Users can facilitate problem-solving through guided interactions from fine-tuned LLM, ensuring that aimless queries are supported in a coherent and purposeful manner.

\subsection{Application of the Framework}\label{sec: workflow}
\begin{figure}[h]
\vspace{-0.3cm}
  \centering 
  \includegraphics[width=\columnwidth]{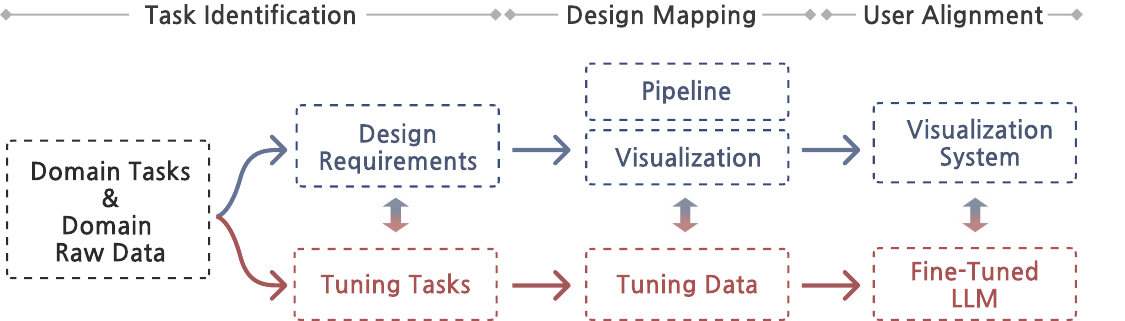}
  \caption{Workflow for applying the framework. All three phases of the workflow are designed to achieve the alignment challenges.}
  \label{fig: workflow}
  \vspace{-0.5cm}
\end{figure}

% Our conceptual framework analyzes the relationships between domain tasks and domain data, humans, visualization systems, and fine-tuned LLMs. 
We can derive insights for implementing intelligent visualization interaction processes from the conceptual framework.
% 因此，我们提出了一个general workflow以及对应的guidelines，并根据在一个教育问题上的实践进行详细解释。我们针对SRL问题展开， 
% workflow由以下三个阶段组成，我们通过箭头的方式来表示为满足对齐而产生的作用与影响：
\textcolor{revision}{Therefore, we propose a general workflow (Fig.~\ref{fig: workflow}) along with corresponding guidelines and provide a detailed explanation based on its application to an educational problem. We focus on SRL, a process where students plan, monitor, and evaluate their own learning, as intelligent and online learning platforms have shifted students toward greater self-initiative rather than traditional teacher guidance.}
\textcolor{revision}{The workflow process undergoes three phases, with guidelines using ``$\Rightarrow$'' to represent the influence or guidance of one aspect on another.}

% Therefore, in Sec.~\ref{sec: general workflow}, we summarize the general workflow \textcolor{revision}{with guidelines} and analyze the application of workflow to SRL in education (Sec.~\ref{sec: application in education}).

% \subsubsection{General Workflow \textcolor{revision}{\& Guidelines}} \label{sec: general workflow}
% Considering the impact of fine-tuned LLM incorporation and the three alignment challenges mentioned in Sec.~\ref{sec: problem definition}, we propose the workflow shown in Fig.~\ref{fig: workflow}. Beginning with domain tasks and data, we aim to implement an intelligent interactive visualization system supported by the fine-tuned LLM.
% The workflow process undergoes the following three phases:
 
\textbf{Task Identification.}
% system
% 在传统的可视化系统流程中，首先要做的根据领域专家与目标用户的需求提炼设计需求。在这个工作流中也不例外，从领域任务和收集的数据出发进一步明确对可视化系统的需求。
\textcolor{revision}{In a traditional visualization workflow, the first step is to refine \textit{design requirements} based on the needs of domain experts and target users.}
% To align LLMs with domain problems (\ref{A1}), it is important to obtain the needs from domain experts and target users and summarize them as \textit{design requirements}.
% model
% 基于可以支持的用户需求以及现有LLM对领域能力领域知识掌握的不足，我们从中引出tuning tasks以提高模型知识和行为。
\textcolor{revision}{Based on the supported user needs and the existing LLM's gaps in domain knowledge and capabilities (\ref{A1}), we derive targeted \textit{tuning tasks} to enhance the model's domain knowledge and performance.}
% Tasks that require intelligent capability enhancement are extracted from the design requirements for fine-tuning. 
% The \textit{tuning tasks} need to be considered to improve domain knowledge and behaviors.
\begin{itemize}[parsep=0.1pt]
    \item \textcolor{revision}{(Design Requirements $\Rightarrow$ Tuning Tasks)~Extract tasks that require enhanced intelligence from the requirements.}
    \item \textcolor{revision}{(Tuning Tasks $\Rightarrow$ Design Requirements)~Constructing usage scenarios helps to validate and iterate on the needs of potential users.}
\end{itemize}

% 教育领域上的运用
% 在教育领域的SRL问题上，我们从一个preliminary study获取帮助初学者深入理解知识与个性化学习过程设计需求。并从中提炼出优化领域知识问答与个性化推荐等智能决策任务作为微调的主要任务。同时，在具体化功能的同时我们反复与用户进行交流与对接，对智能科学的SRL过程进行了更详细的补充。

\textcolor{revision}{Regarding its application in education, we conducted a preliminary study on SRL to identify the design requirements for helping beginners deeply understand knowledge and personalize their learning process. From this, we distilled key tasks for fine-tuning, such as optimizing domain-specific Q\&A to enhance knowledge comprehension and providing personalized recommendations to tailor the learning experience. We refined these functionalities through continuous user engagement to enhance the intelligent SRL process.}

\textbf{Design Mapping.}
% system
% 基于Design requirements,我们进一步设计可视化视图，以及可视化交互探索流程（visual exploration pipeline)的总结，这是一个正常的可视化流程。
\textcolor{revision}{Based on the design requirements, we further developed \textit{visualization} views and summarized the visual exploration \textit{pipeline}.}
% Based on the design requirements, we need to design the \textit{pipeline} to summarize the domain task processing flow. At the same time, data features and insights are used for \textit{visualization} design (\ref{A2}).
% model
% 融合LLMs让整个可视化过程更加智能与可交互性，因此我们需要收集并构建对应tuning data来适应可视化系统。
\textcolor{revision}{By integrating LLMs, the entire visualization process becomes more intelligent and interactive. We need to collect and construct \textit{tuning data} to adapt to the visualization system (\ref{A2}).}
% The \textit{tuning data} is designed to enhance the model's domain expertise and needs to be aligned with visualization aspects. 
% This alignment is reflected in the visualization design, data features, and the context of data insights.

\begin{itemize}[parsep=0.1pt]
    % pipeline->data: 我们需要总结可视化交互解决问题的过程中的pattern，以多轮对话或思维链的方式指导模型行为。
    \item \textcolor{revision}{(Pipeline $\Rightarrow$ Tuning Data)~We need to summarize patterns in the process of solving problems through visual interactions, guiding model behavior using multi-turn dialogues or CoT approaches.}
    % \item \textcolor{revision}{(Pipeline $\Rightarrow$ Tuning Data)~The pipeline presents the logic flow of problem processing, assisting in the construction of multi-turn dialogue data or thought chains.}
    % visualization->data: 模型需要支持不同可视化视图中的功能，这需要构建符合可视化视图设计的数据形式，包括理解编码方式、数据来源特征以及可视化表示的数据洞察形式。例如对知识点的归纳，是结合了节点链接图结构而构建的知识图谱，以进行多个关联知识点的推荐的。同时，这也一定程度规范了数据构造的结构与指令格式。
    \item \textcolor{revision}{(Visualization $\Rightarrow$ Tuning Data)~To support the functionalities of different visualization views, we need to construct data structures that align with their designs. This includes understanding encoding methods, data source characteristics, and forms of data insights. 
    % 把所有举例的、和教育相关的放到后面去说了
    % For instance, in summarizing knowledge points, we use a node-link graph structure to build a knowledge graph, which helps in recommending related knowledge points and enhances the model's understanding of the relationships between them. 
    Additionally, the mapping relationships in visualization standardize the data structure and instruction format.}
    % \item \textcolor{revision}{(Visualization $\Rightarrow$ Tuning Data)~The data formats behind visual designs standardize the structural consistency of model outputs. Data features and insights of visualizations offer varied tuning data types, such as correlation insights and network structures for relationship extraction, and change points or trends for identifying key entities.}
    % data->pipeline: tuning data对pipeline的设计进行补充和优化，将智能化交互融合到可视化系统的探索过程中。
    \item \textcolor{revision}{(Tuning Data $\Rightarrow$ Pipeline)~Concrete tuning data supplements and optimizes the pipeline design, integrating intelligent interaction into the exploration process.}
    % \item \textcolor{revision}{(Tuning Data $\Rightarrow$ Pipeline)~Concrete data supplements, specify and refine the pipeline process.}
    % data->visualization: 根据微调数据，模型生成符合预期的可视化表示的数据，提高了可视化结果的准确性。通过对数据的深入分析，更多的可视化设计选择丰富了数据的可视化表现。
    \item \textcolor{revision}{(Tuning Data $\Rightarrow$ Visualization)~Based on the tuning data, the model generates data that aligns with the expected visualization views, improving the accuracy of the results. In-depth data analysis provides more visualization design options, enriching the representation of the data.}
    % \item \textcolor{revision}{(Tuning Data $\Rightarrow$ Visualization)~Data formats and dimensions enhance visualization capabilities, offering more design choices.}
\end{itemize}

% 教育领域上的运用
% 在自学问题中，根据将SRL pipeline过程细分为了forethought、performance以及self-reflection的详细自学流程，结合微调任务，我们进行了4个场景数据的构造。例如，为了符合self-reflection阶段自测的交互流程，我们以问题推荐-解答-讲解的多轮对话形式构建微调数据。同时，我们为不同可视化视图构建支持的指令对话数据，例如对知识点的归纳，是结合了节点链接图结构而构建的知识图谱，以进行多个关联知识点的推荐的。同时，这也为这个节点链接网络图提供了关系引导、自动添加的功能。
\textcolor{revision}{In the context of SRL, based on the detailed pipeline process segmented into forethought, performance, and self-reflection phases, we constructed data for four scenarios in alignment with fine-tuning tasks. For example, to align with the interactive process of self-assessment in the last phase, we constructed fine-tuning data in the form of multi-turn dialogues encompassing question recommendations, answers, and explanations. Additionally, we developed supported instruction fine-tuning data for various visualization views. For instance, summarizing knowledge points involved constructing a knowledge graph with a node-link network view to facilitate the recommendation of multiple related knowledge points. This also provided relational guidance and automatic addition functionalities for the node-link network view.}

\textbf{User Alignment.}
% system
A prototype of the \textit{visualization} is used to obtain user suggestions for iterative optimization to capture user intent (\ref{A3}).
% model
From user feedback, tuning data should be refined to improve the performance of \textit{fine-tuned LLM}. 
% In order to satisfy random and varied user queries, the model also needs to implement guided outputs of insights or task realizations.

\begin{itemize}[parsep=0.1pt]
    % system->model: 系统的用户体验优化了模型的推荐能力，增强了交互式推荐和意图推断。从交互记录中发现用户特征和交互模式，有针对和个性化的改进模型输出。
    \item \textcolor{revision}{(Visualization System $\Rightarrow$ Fine-tuned LLM)~The system's user experience optimizes the model's recommendation capabilities, enhancing interactive recommendations and intent inferences. By analyzing interaction records, we identify user characteristics and interaction patterns, enabling targeted and personalized improvements to the model's output.}
    % \item \textcolor{revision}{(Visualization System $\Rightarrow$ Fine-tuned LLM)~The system's user experience optimizes recommendation tasks for the model, enhancing interactive recommendations and intent inference capabilities.}
    % \item \textcolor{revision}{(Visualization System $\Rightarrow$ Fine-tuned LLM)~Customize model outputs based on user characteristics and specific needs.}
    % model->system: 将微调之后的模型接入系统每个模块，提供更多智能交互的方式支持最初的领域问题。
    \item \textcolor{revision}{(Fine-tuned LLM $\Rightarrow$ Visualization System)~Integrating the fine-tuned model into each system module provides more intelligent interaction methods to support the initial domain problem.}
    % \item \textcolor{revision}{(Fine-tuned LLM $\Rightarrow$ Visualization System)~Intelligent outputs offer more interactive options, enhancing user engagement.}
\end{itemize}

% 教育领域上的运用
% 在教育中，将微调之后的模型和可视化系统相结合组成了最终的系统Tailor-Mind，用于原型模型收集用户体验。根据这些反馈，我们将用户群体（学生）特征编码给模型，优化模型能力。例如，根据初学者这种用户特点，提供提问推荐以简单化提问过程，发散思维。基于模型回答看作引用材料，进一步智能化提取关键信息与结构化数据，帮助学生掌握重点。
\textcolor{revision}{In education, we combined the fine-tuned model with the visualization system to form the final system, Tailor-Mind, for prototype user experience collection. Based on user feedback, we encoded user group (students) characteristics into the model to optimize its capabilities. For instance, considering beginners' traits, we provided question recommendations to simplify the questioning process and encourage divergent thinking. By treating model answers as reference material, we further tuned the model to intelligently extract key information and structured data, helping students grasp essential points.}

\section{Requirement Analysis of Tailor-Mind} \label{sec: requirement analysis}
\textcolor{revision}{
% The emergence of intelligent and online learning platforms has catalyzed a shift towards SRL, where students increasingly rely on their own initiatives rather than traditional teacher-guided methods. 
% SRL involves learners actively controlling and monitoring their learning processes, including setting goals, selecting strategies, and evaluating progress.
Recognizing the need for effective tools to support SRL,} we conducted a preliminary study to identify the design requirements of the visualization system and specific fine-tuning tasks for intelligent SRL.
%% Preliminary study
\subsection{Interviews and Surveys} \label{sec: interviews and surveys}
We interviewed two domain experts and investigated how students adopt SRL in their studies.

% Expert interview
Our research involved two thirty-minute interviews with a distinguished education researcher (E1), who studies students' learning behaviors, and a university lecturer (E2) who teaches AI. E1 advocated for Zimmerman's SRL model~\cite{zimmerman2000srl} to guide our study, emphasizing the need to cultivate students' initiative and enthusiasm. E2, from a teaching perspective, noted the difficulty students face in linking various knowledge areas. E2 suggested that an ideal ITS should reduce cognitive load, enhance learner engagement, and offer personalized knowledge representation. Both experts concurred on the significance of promoting SRL over passive or task-specific learning methods.

\begin{figure}[h]
\vspace{-0.2cm}
  \centering 
  \includegraphics[width=\columnwidth]{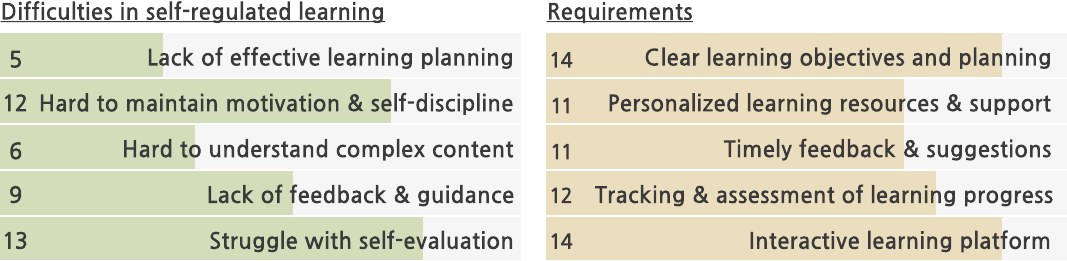}
  \caption{Results of student surveys. The left chart indicates the difficulties encountered by beginners in SRL. The right side demonstrates their need for a visualization system that supports intelligent aids to SRL.
}
  \label{fig: preliminary study}
  \vspace{-0.35cm}
\end{figure}

% Student survey
We also recruited students with experience in machine learning and deep learning courses and received 16 responses (7 female, 9 male). This group included 3 Ph.D. students, 8 M.S. students, and 5 undergraduates, all from the Computer Science and Data Science disciplines. The results showed that all participants had attempted SRL but were confused about practical implementations. Their main challenges and needs are depicted in Fig.~\ref{fig: preliminary study}. Notably, almost everyone desired personalized and timely assessments of their learning outcomes.

%% Challenges
\subsection{Challenges for Self-Regulated Learning}\label{sec: challenge}
Through interviews with experts and student surveys, we identified several main challenges during the SRL process.
\begin{enumerate}[label=C\arabic*,itemsep=0pt, parsep=0pt]
\item \label{C1} \textbf{Limited knowledge of SRL.} E1 indicated that the primary challenge for many students in engaging with self-regulated scientific and effective learning stems from a lack of awareness. This observation aligns with findings from our student surveys.
\item \label{C2} \textbf{Lack of motivation and guidance.} E1 mentioned that maintaining enthusiasm and focus is difficult, especially in self-learning. She emphasized that appropriate goal-setting and guidance are crucial for sustaining students' self-motivation and self-discipline throughout the typical SRL process.
\item \label{C3} \textbf{Complex and esoteric knowledge.} E2 highlighted that a major barrier for many students is the complex organization of knowledge. It is a great challenge to understand, apply, and interconnect different concepts independently. This complexity often overwhelms students, hindering their ability to achieve satisfactory learning outcomes through SRL.
\item \label{C4} \textbf{Lack of immediate feedback.} The cost and accessibility of personalized tutoring present significant barriers. Existing methods fall short of providing personalized feedback as well. As shown in Fig.~\ref{fig: preliminary study}, assessing their learning progress is challenging.
\end{enumerate}

%% Design requirements
\subsection{Requirements for Tailor-Mind}\label{sec: requirements}
In response to the challenges outlined in Sec.~\ref{sec: challenge}, we present the following requirements for the intelligent assistance of SRL. 

\begin{enumerate}[label=R\arabic*,itemsep=0pt, parsep=0pt]
\item \label{R1} \textbf{Comprehensive SRL guidance and awareness building.} Tailor-Mind should facilitate users in adhering to the SRL process (\ref{C1}) to foster active learning. Leveraging insights from educational models and theories, it's critical to specify and encode detailed sub-tasks within the foundational SRL framework. We should also educate users about the significance of setting goals, implementing effective learning strategies, and reflecting (\ref{C2}).
\item \label{R2} \textbf{Optimization of learning depth and efficiency.} \textcolor{revision}{We must align our teaching goals to provide clear and well-organized explanations.} To support beginners in applying and transferring knowledge while reducing cognitive load, presenting knowledge in a structured and simplified way is crucial (\ref{C3}). Visual representations are important in breaking down complex information and illustrating the relationships between knowledge points.
\item \label{R3} \textbf{Personalized learning and adaptive assessments.} Tailoring the learning journey to individual needs is crucial. Offering users intelligent recommendations for learning objectives, paths, and content can significantly enhance personalized learning experiences. By analyzing learning performance across various objectives, Tailor-Mind must deliver targeted and suitable feedback to facilitate dynamic and iterative learning, empowering students to progress effectively and adaptively (\ref{C4}).
\item \label{R4} \textbf{Engaging and interactive learning environments.} To keep students' enthusiasm and focus during SRL (\ref{C2}), Tailor-Mind should provide visual guidance throughout the learning process via interactions. Simultaneously, the explanation of intricate knowledge points should be captivating and engaging (\ref{C3}).
\end{enumerate}
\section{Tailor-Mind} \label{sec: model & system}
\begin{figure*}[htbp]
% \vspace{-0.4cm}
  \centering 
  \includegraphics[width=\linewidth]{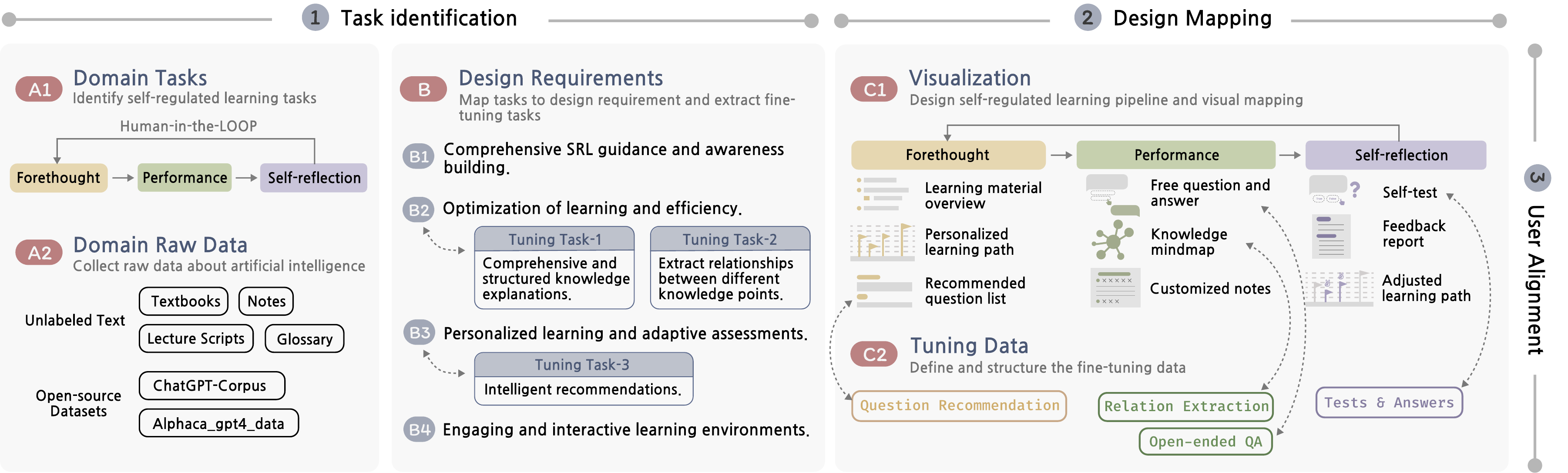}
  \caption{In applying workflow to SRL in education, we outline the process in three phases. Phase \raisebox{-0.1em}{\includegraphics[height=1em]{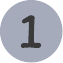}} involves establishing a fundamental understanding of the SRL task (A1) and collecting data on artificial intelligence (A2). The design requirements (B) align with those outlined in Sec.\ref{sec: requirements} from which we derive the tuning tasks. Phase \raisebox{-0.1em}{\includegraphics[height=1em]{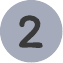}} details the SRL pipeline sub-tasks and visualizations (C1), leading to the creation of fine-tuning data (C2). In phase \raisebox{-0.1em}{\includegraphics[height=1em]{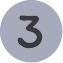}}, we enhance the fine-tuning effects and visualization interactions by integrating user feedback within the visualization system.}
  \label{fig: tailormind workflow}
  \vspace{-0.4cm}
\end{figure*}

In this section, we follow the workflow (Fig.~\ref{fig: workflow}) of the proposed conceptual framework in Sec.~\ref{sec: framework} to facilitate the SRL process, as shown in Fig.~\ref{fig: tailormind workflow}. Specifically, Sec.~\ref{sec: SRL pipeline} details the learning process and identifies sub-tasks under each stage. In Sec.~\ref{sec: fine-tuned LLM}, we introduce the fine-tuning tasks and corresponding tuning datasets to align the domain requirements with LLMs (\ref{A1}). We purposely analyze some forms of data construction that consider visualization (\ref{A2}) and user interaction alignment (\ref{A3}). The user interface of Tailor-Mind is illustrated in Sec.~\ref{sec: user interface}.

\subsection{Workflow and Self-Regulated Learning Process}  \label{sec: SRL pipeline}
Guided by the conceptual framework proposed in Fig.~\ref{fig: framework}, we have integrated the fine-tuned LLMs into a visualization system that supports intelligent SRL for beginners. The specific implementation workflow is shown in Fig.~\ref{fig: tailormind workflow}. Starting with the domain tasks (Fig.~\ref{fig: tailormind workflow}A1), we conducted a detailed requirement analysis (Fig.~\ref{fig: tailormind workflow}B). To help AI beginners with a comprehensive SRL process (\ref{R1}), we propose a detailed SRL pipeline (Fig.~\ref{fig: tailormind workflow}C1) based on Zimmerman's model~\cite{zimmerman2000srl}. The sub-tasks in the pipeline serve as the main objectives for visualization design and fine-tuning work. Finely segmenting the domain space aids in better aligning the model with domain problems. The pipeline consists of three stages: forethought, performance, and self-reflection.

% forethought
Forethought involves planning and goal-setting. Analyzing user-uploaded learning materials, we recommend personalized learning paths (\ref{R3}), helping beginners organize resources and set achievable goals, thereby enhancing motivation (\ref{R4}).
% performance
During the performance stage, beginners employ strategies from the forethought stage. Besides acquiring knowledge from the LLM, we facilitate the application of this knowledge, for instance, through notepads and knowledge mind-maps, enabling learners to track their progress.
% self-reflection
Self-reflection, identified as the most challenging stage for beginners, involves synthesizing learning into structured notes. Based on evaluations from tests aligned with set goals, we provide feedback and help students dynamically adjust learning paths for continuous learning experiences (\ref{R3}).

\subsection{Fine-tuned LLM for Self-Regulated Learning}\label{sec: fine-tuned LLM}
Through the task identification in Fig.~\ref{fig: tailormind workflow}-\raisebox{-0.1em}{\includegraphics[height=0.9em]{figs/tailormind-workflow-phase1.pdf}}, we have collected raw data for AI teaching (Fig.~\ref{fig: tailormind workflow}A2), including unlabeled data and conversation data related to the topic. 
% 把第一次迭代过程也讲一下，说明前后两次在任务设定上的区别
% 我们对微调任务以及具体数据形式的设定进行了一次迭代，这是由于我们将原型模型和系统收集用户反馈之后的调整。初次微调任务只关注领域知识的增加，即帮助知识增强与扩展，这无法满足可视化与用户交互对齐。因此我们进一步基于设计需求和粗数据细化了微调任务，以及设定了四个场景。
\textcolor{revision}{We iterated on the fine-tuning tasks and the specific data formats, prompted by adjustments made after collecting user feedback with the prototype model and system. The initial fine-tuning tasks focused solely on augmenting domain knowledge, aimed at enhancing and expanding knowledge, which did not incorporate visualization design mapping and user alignment. Consequently, we refined the tuning tasks (T1-T3) based on design requirements and raw data~(Fig.~\ref{fig: tailormind workflow}B) and established four scenarios.} 

\begin{figure}[h]
% \vspace{-0.2cm}
  \centering 
  \includegraphics[width=\columnwidth]{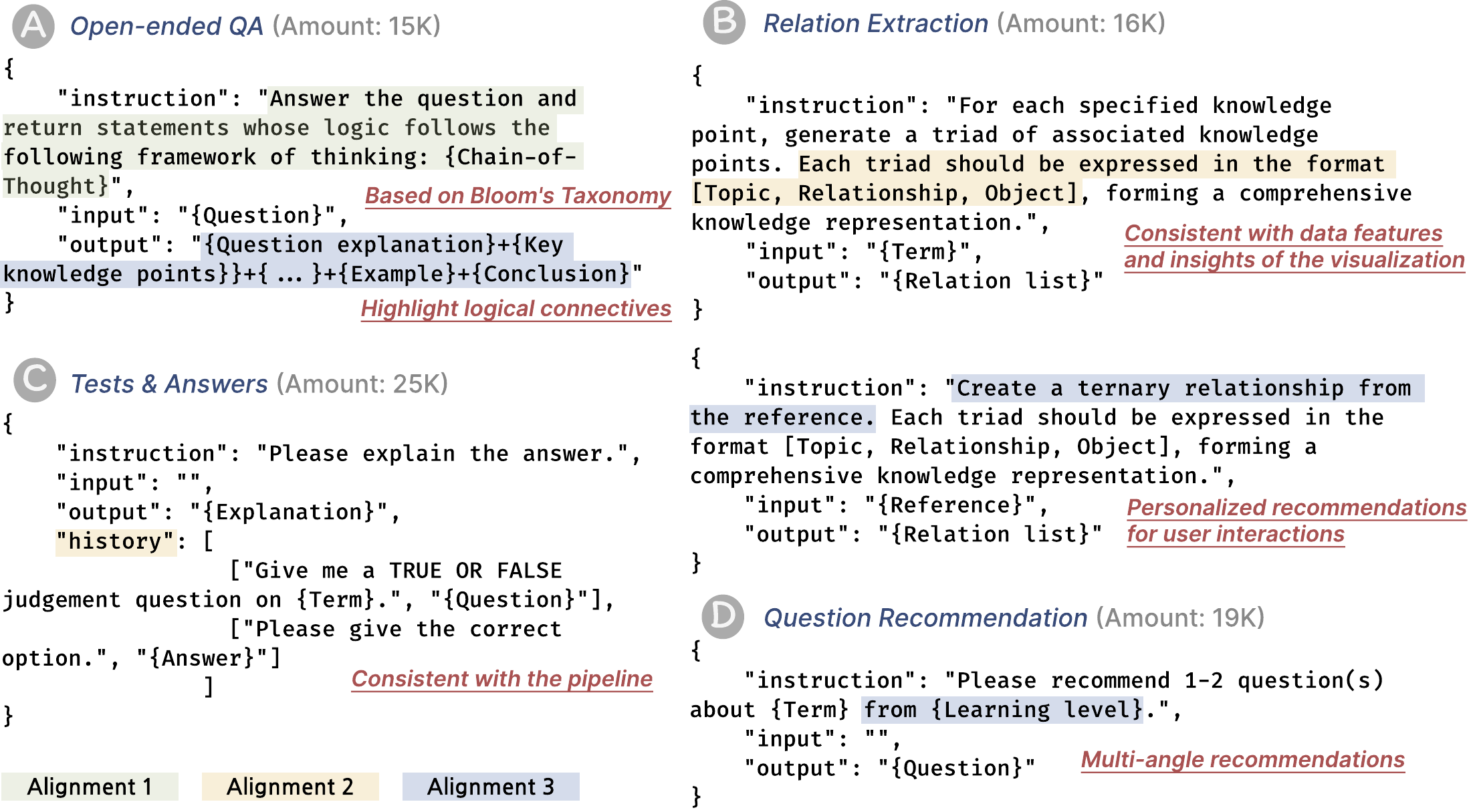}
  \caption{\textcolor{revision}{Refined fine-tuning datasets with examples, where different highlights indicate various alignments. Data construction for tuning tasks~(A for T1, B for T2, C and D for T3) has undergone one iteration.}}
  \label{fig: datasets}
  \vspace{-0.6cm}
\end{figure}

\textcolor{revision}{\textbf{T1 Comprehensive and structured knowledge explanations.} Initially, multi-turn dialogues were extracted from raw data. Focusing on learning efficiency and cognitive depth, we integrated Bloom's Taxonomy~\cite{david2002bloom} into the model's thought chain~\cite{Wei2022ChainOT} in \textit{Open-ended QA}~(Fig.~\ref{fig: datasets}A). User feedback analysis has led to incorporating visual suggestions, such as highlighting, in the responses.}

\textcolor{revision}{\textbf{T2 Extract relationships between different knowledge points.}} The \textit{Relation Extraction} reflects the networked structural data of disciplinary knowledge~(Fig.~\ref{fig: datasets}B). \textcolor{revision}{After the initial iteration, to assist users in linking paragraph text to network visualization, we expanded this task to arbitrary texts, aiding users in obtaining personalized relationship recommendations based on model responses as references.} Fine-tuning for data features helps bridge the gap between model output and data visualization, thus enhancing user comprehension and interaction.

\textcolor{revision}{\textbf{T3 Intelligent recommendations.} Intelligent recommendations primarily involve suggesting \textit{Tests \& Answers} and \textit{Question Recommendation}. In Fig.~\ref{fig: datasets}C, aligned with the self-reflection stage of the pipeline, we utilized a multi-turn dialogue format. This strategy provides the model with a contextual environment conducive to timely feedback, incorporating phases of question recommendation, answering, and detailed explanations.} Through the visualization system, the model recommends questions at different learning levels~(Fig.~\ref{fig: datasets}D), guiding beginners in a structured and comprehensive learning approach.

Based on the described tasks and scenarios, we have compiled a total of 74,932 fine-tuning data entries. The construction of tuning data involves extracting multi-turn dialogues from reference~\cite{Yang2023RefGPTDG} and invoking terms. The fine-tuning process is developed on the open-source LLM Baichuan2-7B-chat~\cite{Yang2023Baichuan2O}, which is trained on a high-quality corpus of 2.6 trillion tokens, achieving a high level of performance in both English and Chinese. We conducted supervised low-rank adaptation~\cite{hu2022lora} fine-tuning on the constructed tuning dataset, which endows the model with knowledge reasoning and educational behavior patterns. The training process had a learning rate of 5e-5, underwent 3 training epochs, and was completed on 4×4090 GPUs. We describe the entire data construction and fine-tuning process in Supplementary Materials.

\subsection{User Interface}\label{sec: user interface}
We design a visual interface including the views and interactions representative of the SRL pipeline outlined in Sec.~\ref{sec: SRL pipeline}. In Fig.~\ref{fig: tailormind workflow}-\raisebox{-0.1em}{\includegraphics[height=0.9em]{figs/tailormind-workflow-phase3.pdf}}, we consider the alignment of the user interaction with the fine-tuned model.

\subsubsection{Chat View}
The Chat View (Fig.~\ref{fig:teaser}A) serves as a primary gateway, supporting the upload of learning materials and providing a guided introduction to SRL. We consider the presentation of LLM outputs in the Chat View as a visual representation of knowledge, summarizing the multi-turn dialogues between users and LLMs as interactions. \textcolor{revision}{Therefore, to reduce cognitive load, we will emphasize logical connectives in responses such as \textit{For instance} (Fig.~\ref{fig:teaser}A2). In general, the responses will be carried out according to the following steps: interpreting the question, explaining key knowledge points, giving examples, and summarizing.} The visualization of knowledge extends beyond text dialogues to include rich text formats, such as button selections (Fig.~\ref{fig:teaser}A1). As illustrated in Fig.~\ref{fig:teaser}A3, selecting some buttons about knowledge relationships can trigger the addition of new nodes and edges in the Knowledge MindMap.

\subsubsection{File Preview}
The File Preview (Fig.~\ref{fig:teaser}B) provides users with a preview of the uploaded learning materials (Fig.~\ref{fig:teaser}B1). Given the fine-tuned model's limitations in multimodal processing, ChatGPT is employed to parse the learning materials, thereby ensuring the precision of the subsequent outcomes. The knowledge structure tree widget (Fig.~\ref{fig:teaser}B2) is constructed based on the relationships of knowledge points within the file. Users can interact through clicks to view the corresponding knowledge cards (Fig.~\ref{fig:teaser}B3). These cards shed light on the significance or application of the respective knowledge points in the context of the material. Each knowledge card features the ability to copy and ask questions, which facilitates a seamless transition to the Chat View for in-depth explanation and reduces aimless exploration for beginners.

\subsubsection{Question Recommendation}
Integrating Bloom's Taxonomy with educational objectives, we classify eight learning levels from basic to advanced. The fine-tuned model recommends questions based on the knowledge points extracted from the file across the eight learning levels. Additionally, each question supports content copying and resetting (Fig.~\ref{fig:teaser}E).

\subsubsection{Learning Path}
We depict key knowledge points along the learning path as milestones, visualized as small flags in Fig.~\ref{fig:teaser}F. Each flag's color corresponds to the learning level required for that knowledge point, and flag height represents the importance. Larger flags signify greater significance. Beneath each flag, specific expressions of knowledge are encoded as colored dots, with colors matching those used system-wide for learning levels and sizes denoting importance. Hovering over these flags reveals detailed expressions derived from the model. Numerical analysis is facilitated by stacked bar charts that tally these colored dots, allowing for comparison of learning paths before and after the self-reflection stage. The timeline on which these milestones are placed uses a relative scale to represent time spent between them, as we lack direct access to the users' actual learning abilities.

Incorporating the importance and relevance of knowledge points, the system offers personalized recommendations. Users can customize their learning path within the Knowledge MindMap using reset, edit, and submit buttons to adjust milestone data as needed.

\subsubsection{Knowledge MindMap}
\textcolor{revision}{A knowledge point is a fundamental concept or piece of information that serves as a building block, enabling students to incrementally understand complex AI systems.} Based on this, we represent knowledge point entities as nodes and illustrate the logical relationships between them as edges. The nodes’ color and size mirror the design mappings used for milestone flags in the Learning Path. At the same time, tutors suggest that we correspond the relationships between knowledge points to learning levels, ensuring logical consistency and reducing cognitive load for users. To display the structural features of knowledge points, we support various layouts for the network structure (Fig.~\ref{fig:teaser}C1), including the ``dagre'' layout to show hierarchical relationships and the ``concentric'' layout to highlight core knowledge points. Each node supports recommended questions, setting goals, and taking notes (Fig.~\ref{fig:teaser}C3). Questions recommended regarding this knowledge point can be discussed directly in Chat View (Fig.~\ref{fig:teaser}A1). After completing the note-taking, the model processes it into a word cloud returned on the selected node (Fig.~\ref{fig:teaser}C2), which is convenient for preview and can serve as a learning marker. Users can also customize the network structure based on their understanding of knowledge points, performing operations such as adding, deleting, and editing nodes or edges. In addition to taking notes on a particular knowledge point node, users can also record any discoveries in Fig.~\ref{fig:teaser}D.
\section{Evaluation} \label{sec:evaluation}
In this section, we evaluate the fine-tuned LLM (SFT-2.0) by comparing the performances of the other four models on the test datasets through both human and OpenAI GPT-4~\cite{openai2024gpt4} (Sec.~\ref{sec: model performance}). Two usage scenarios are narrated to illustrate how Tailor-Mind can help the SRL process in Sec.~\ref{sec: usage scenario}. We further conduct an in-person study and interviews with participants and analyze the results in Sec.~\ref{sec: user study}.

\subsection{Model Performance} \label{sec: model performance}
%% Why don't we use the objective evaluation datasets?
% no benchmark datasets in AI education + general benchmark has no meaning
There is a lack of authoritative benchmark datasets for evaluation in AI education. The generic benchmark dataset, such as AGIEval~\cite{zhong2023agieval} and C-Eval~\cite{huang2023ceval}, is intended to general models and is not suitable to test a specific model's output with a standardized structure. Therefore, we consider constructing a dedicated test dataset, setting subjective evaluation criteria, and comparing the performances among several models. The evaluation results are shown in Table.~\ref{tab:performance}. Detailed evaluation results and analyses are available in the Supplementary Materials.

\textbf{Settings.}
%% test datasets
% Construction + information
% 我们使用创建特定于AI8个子领域下对应的7个微调任务设定测试集问题，并为每个问题提供最优答案。这种设计能够确保LLMs在理解和处理领域具体问题时的准确性和深度，反映出模型在吸收和应用领域知识上的能力(A1)。
% 为确保模型输出的可视化表示与设计目的、数据特征和领域洞察的一致性(A2)，我们在评估过程中特别强调输出的准确性、完整性和清晰性。这不仅适用于语言输出，也至关重要于可视化输出。
% 在评估设计中引入了专家和多轮评分机制，模仿了真实世界中用户与系统的交互模式，包括对模型反馈的不确定性和探索性特征。这种方法有助于评估模型在理解用户意图和行为背后的逻辑上的表现，进而提升模型对用户交互的对齐能力(A3)。
\textcolor{revision}{We generated a dataset comprising 280 test data entries for seven fine-tuning tasks across eight AI subdomains using GPT-4, aiming to cover a wide range of issues within the domain as comprehensively as possible. Each task included five questions of varying difficulties, each accompanied by an optimal answer for subsequent evaluation reference. This dataset assesses LLMs' accuracy and depth in understanding and addressing domain-specific issues, thereby reflecting the models' ability to assimilate and apply domain knowledge (\ref{A1}).} Throughout this process, we conducted manual reviews and engaged in self-reflection with GPT-4 to ensure the dataset's accuracy\cite{ji2023towards}. 
%% different models
We selected the Base-model, EduChat~\cite{dan2023educhat}, OpenAI GPT-3.5~\cite{openai2023chatgpt} and the SFT-1.0 model (without user alignment and visualization alignment), to compare with the final fine-tuned model SFT-2.0. \textcolor{revision}{These models were chosen to facilitate a multifaceted comparison, including domain expertise, output consistency, and alignment requirements, as detailed in the Supplementary Material.} To ensure fairness and objectivity, model information was kept undisclosed to referees.

\textbf{Methods.}
% Human evaluation
% GPT-4 evaluation
% 3 metrics + reference 
\textcolor{revision}{To assess model performances and their alignment with user perceptions (\ref{A3}), we introduced a referee model to simulate real-world scenarios.} Leveraging GPT-4, known for its alignment with controlled and crowdsourced human preferences~\cite{Zheng2023JudgingLW}, we employed it to evaluate outputs based on \textit{Accuracy}, \textit{Completeness}, and \textit{Clarity}, scoring each criterion from 0 to 5. \textcolor{revision}{These criteria facilitate the evaluation of model outputs for consistency with design intentions, data characteristics, and domain insights (\ref{A2}), with interpretations varying slightly across different tasks.} Specifically, \textit{Accuracy} assesses alignment with the reference answer in content, semantics, and structure, \textit{Completeness} ensures no detail is overlooked, and \textit{Clarity} evaluates logical coherence and clear expression. To mitigate bias, ground truth is provided. \textcolor{revision}{Scores are averaged over multiple rounds to capture different dimensions and simulate user interactions, addressing uncertainties and the exploratory nature of model feedback (\ref{A3}).} Additionally, seven AI experts manually rated these criteria to enrich the evaluation process.

\textbf{Results.}
%% Summary
From Table.~\ref{tab:performance}, the SFT-2.0 model outperforms the other models in all aspects of the evaluation by humans. The following findings are drawn from the results:
%% ACC + CLR
(1) The SFT-2.0 model's responses are accurate and follow the logic of knowledge presentation. Experts generally indicated that the model output was highly structured and could be aligned with the subsequent visualization design. However, other models, even when given a detailed prompt, still did not fulfill all the requirements.
%% std
(2) The SFT-2.0 model exhibits the most stable performance with the smallest variance across the three criteria, primarily due to the benefits and effects of fine-tuning. The stability of model outputs is particularly important for user interaction and presentation in the visualization system.
%% CPL
(3) The SFT-2.0 model is more in line with user preferences. Although we emphasized that the referee model should not be influenced by response length when scoring completeness, it still tended to judge longer responses as better. This issue particularly existed in the Question Recommendation task, leading to the result in Table.~\ref{tab:performance} that considered GPT-3.5's answers more complete. Experts corrected this by pointing out that \textit{"GPT-3.5's answers are redundant and not conducive to direct understanding, and the results from the SFT-2.0 model are more suitable for beginners"}.
\begin{table*}[h]
\centering
\begin{tabular}{|c|c|c|c|c|c|c|c|c|}
\hline
\multirow{2}{*}{Model} & \multicolumn{4}{c|}{Evaluation by Human} & \multicolumn{4}{c|}{Evaluation by GPT-4} \\
\cline{2-9}
 & ACC & CPL & CLR & Average & ACC & CPL & CLR & Average \\
\hline
Base-model & \uline{3.68} ($\pm$1.22) & 3.71 ($\pm$1.22) & 3.28 ($\pm$1.80) & 3.55 & 3.54 ($\pm$1.91) & 3.86 ($\pm$1.77) & 3.75 ($\pm$1.54) & 3.72 \\
EduChat & 3.45 ($\pm$2.33) & 3.42 ($\pm$2.32) & 2.96 ($\pm$2.65) & 3.28 & 3.11 ($\pm$2.98) & 3.45 ($\pm$2.91) & 3.32 ($\pm$2.04) & 3.30 \\
GPT-3.5 & 4.11 ($\pm$0.60) & \uline{3.93} ($\pm$0.58) & \uline{3.79} ($\pm$1.44) & \uline{3.94} & \uline{4.09} ($\pm$1.34) & \textbf{4.09} ($\pm$1.21) & \uline{3.98} ($\pm$0.66) & \uline{4.05} \\
SFT-1.0 & 3.48 ($\pm$1.09) & 3.29 ($\pm$1.17) & 3.22 ($\pm$1.21) & 3.33 & 2.97 ($\pm$2.29) & 3.00 ($\pm$2.03) & 3.35 ($\pm$1.73) & 3.10 \\
SFT-2.0 & \textbf{4.40} ($\pm$0.51) & \textbf{4.03} ($\pm$0.58) & \textbf{4.46} ($\pm$0.54) & \textbf{4.30} & \textbf{4.15} ($\pm$0.98) & \uline{4.06} ($\pm$0.97) & \textbf{4.39} ($\pm$0.58) & \textbf{4.20} \\
\hline
\end{tabular}
\caption{Evaluation of model performance using metrics ACC (Accuracy), CPL (Completeness), and CLR (Clarity), where \textbf{bold} indicates the best result and \underline{underline} the second best. Our model (SFT-2.0) performs well in both human and GPT-4 assessments.}
\label{tab:performance}
\vspace{-0.4cm}
\end{table*}

\subsection{Usage Scenario}\label{sec: usage scenario}
We illustrate usage scenarios with Tailor-Mind from two perspectives: a beginner's enhanced understanding of the Transformer model and a beginner's preparatory journey in Reinforcement Learning (RL).

\subsubsection{Integrating Knowledge and Deepening Understanding}
Evelyn is a data analyst who needs to make sequence predictions for her current work. \textcolor{revision}{Therefore, she uploads the authoritative learning material that she found on a website, hoping to understand the Transformer better and determine whether it meets her work requirements.} In the forethought phase, she discovers that the knowledge points listed in the File Preview (Fig.~\ref{fig:teaser}B2) seemed familiar, but there was significant forgetfulness and a lack of understanding of how they are related to each other. After understanding the file structure and learning path, Evelyn begins her study of model components. By the time she reaches the final milestone, she locates the ``Transformer Encoder'' node in the Knowledge MindMap (Fig.~\ref{fig:teaser}C4) and continues the study based on the recommended questions. She selects the first candidate question (Fig.~\ref{fig:teaser}A1) and intends to understand the encoder's composition against the corresponding part in the material. The clear and structured response in Fig.~\ref{fig:teaser}A2 satisfies her, and she says, \textit{"This example shows me that such a structure represents an encoder block, and it takes multiple encoder blocks to make up an encoder layer"}. 

The multiple relationships suggested at the bottom of the response catch her attention, and she chooses the last button to add to her customized MindMap (Fig.~\ref{fig:teaser}A3) as she has previously learned that the advent of the Transformer replaces many of the scenarios in which Recurrent Neural Networks (RNNs) are used. While editing the added RNNs node, she notices that the only node connected from the ``Transformer Encoder'' is ``Parallel'', which is highly recommended in terms of importance. Therefore, she continues to ask for the recommended questions related to ``Parallel'' and selects the option ``What is parallel processing, and how does it differ from sequential processing?''. After understanding the answer, she says, \textit{"I always knew that the Transformer was superior to RNNs, but I never understood the specific reasons. Now I realize that it's the Transformer's capability for parallelization that makes it better at handling long sequence data, which aligns well with my upcoming work requirements"}. 

In the self-reflection stage, she is asked to answer the question about why Transformer is superior to RNNs (Fig.~\ref{fig: usage scenarios}A). She easily chooses the correct answer and gains a deeper understanding of the point. Moreover, Evelyn expresses that this process allows her to integrate many fragmented pieces of knowledge, enriching her knowledge network.

\subsubsection{Stimulating Interest and Exploratory Learning}
\begin{figure}[h]
\vspace{-0.1cm}
  \centering 
  \includegraphics[width=\columnwidth]{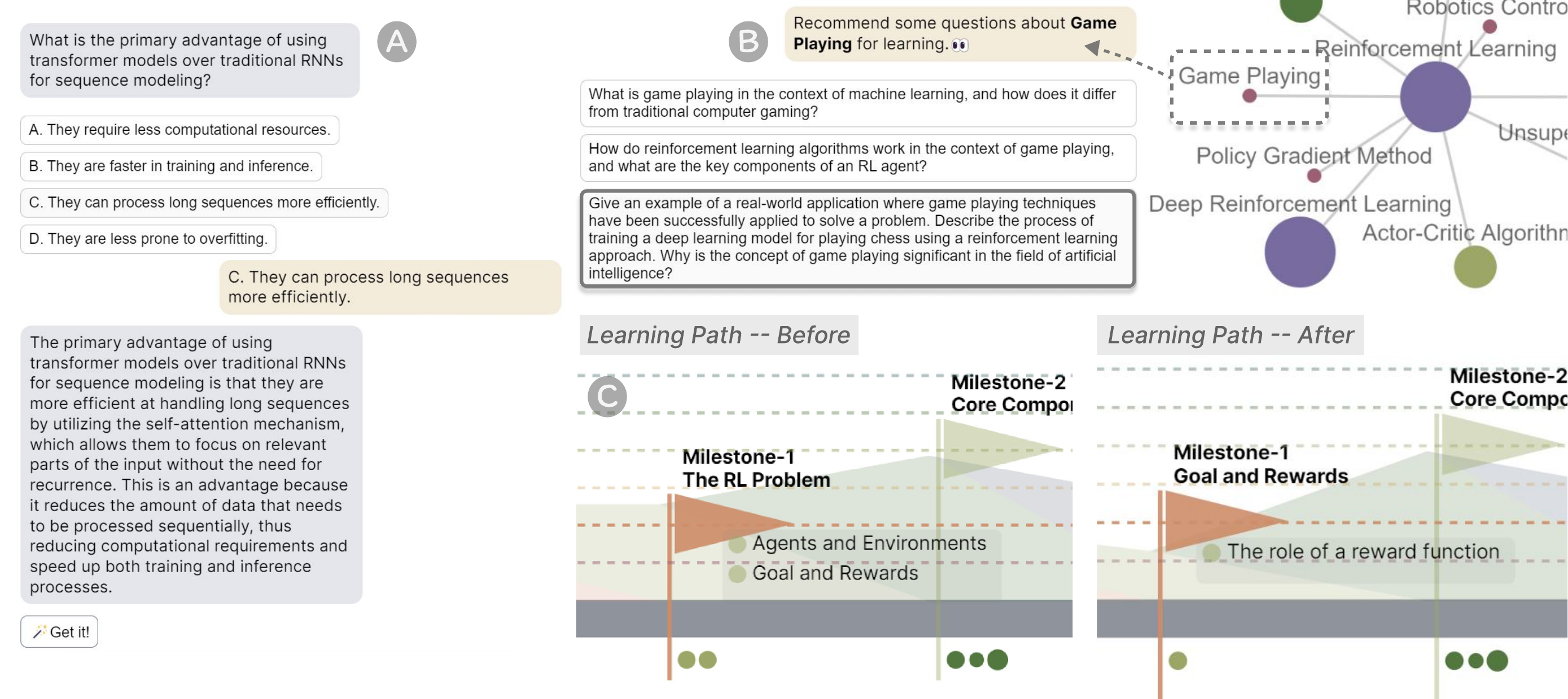}
  \caption{Usage Scenarios for Tailor-Mind. (A) is about the scenario of consolidating understanding. (B) and (C) are intermediate processes in exploratory learning.}
  \label{fig: usage scenarios}
  \vspace{-0.4cm}
\end{figure}

Rex, a senior undergraduate student, is asked to do a preview of the course material about RL. As a result, he seeks the help of Tailor-Mind to sort through the material highlights and lighten his class load. After uploading the material, he follows the recommended learning path for question-driven learning in RL concepts, problems, and core components. While exploring the Knowledge MindMap, he is attracted to the ``Game Playing'' node and triggers the question recommendation function (Fig.~\ref{fig: usage scenarios}B). After understanding the ``Game Playing'' by RL, he becomes more interested in the upcoming course. Rex says, \textit{"Without the intuitive navigation of the RL application provided by Knowledge MindMap, I might regard RL as a somewhat boring topic"}.

Rex's careful and diligent study during the performance stage helps him successfully complete most of the test questions, which filled him with anticipation for upcoming lessons. However, he encounters a mistake due to a vague understanding when answering questions about the ``Reward Function''. The learning path reminds him to reinforce his understanding of the reward function, as shown in Fig.~\ref{fig: usage scenarios}C. Simultaneously, he also brings this question into the classroom.

\subsection{User Study}\label{sec: user study}

We demonstrated the effectiveness of Tailor-Mind through a user study. 
% In light of the challenges listed in Sec.~\ref{sec: challenge}, we assessed the users' learning experience and efficiency separately across the three phases of SRL.
% The learning experience indicates participants' engagement with the comprehensive SRL pipeline, which encompasses establishing clear objectives, enthusiasm and concentration throughout the learning process, feedback on the system's personalized learning design~(note automatic generation and iterative learning mechanism, etc.), and reflection on their learning behaviors.
% Meanwhile, learning efficiency shows the extent of knowledge comprehension, association, application, and learning speed of users. 
We designed a comparative experiment to facilitate learning of the Transformer model, while the control group employed solely GPT-4.
With challenges in Sec.~\ref{sec: challenge} and expert recommendation, we observed 7 metrics of participants' behaviors throughout the process.
They are study duration, the number of questions attempted, question level, study plan adoption, study plan completion rate, note-taking practice, and engagement in self-reflection.
Detailed procedures and experimental records can be found in our Supplementary Materials.

\subsubsection{Experimental Set-up}
\textcolor{revision}{We recruited 24 participants with a background in computer science who have not previously studied the Transformer model. Among them, 8 were undergraduate students, 16 were postgraduate students, and 14 identified themselves as male, 10 as female. We assessed the participants' understanding of SRL and their usual study habits through a pre-study questionnaire and accordingly divided them into two groups, Group T and Group C, with 12 members in each. Group T~(T1-T12) utilized the Tailor-Mind to complete SRL tasks, and Group C~(C1-C12) used the state-of-the-art LLM GPT-4.}

% \subsubsection{Procedure}
We conducted an in-person observational experiment for each participant. The session began with a briefing on the purpose of our user study and an explanation of the whole process.
After a concise 3-5 minute tutorial introducing the participants to SRL's background knowledge, concepts, and procedures, we provided an introduction and tutorial on the visual and interactive features of Tailor-Mind for Group T. An observational study was conducted with all participants to assess their learning experience with specified material on the Transformer model. Throughout this process, the 7 metrics were systematically recorded. \textcolor{revision}{The results of these observational metrics are documented in the Supplementary Material.} 
Upon completing the SRL tasks, participants answered objective questions to assess their learning performance. Additionally, face-to-face interviews were conducted to explore insights and issues observed. Group T participants also responded to supplementary subjective questions regarding their use of Tailor-Mind.

\begin{figure}[h]
\vspace{-0.2cm}
  \centering 
  \includegraphics[width=\columnwidth]{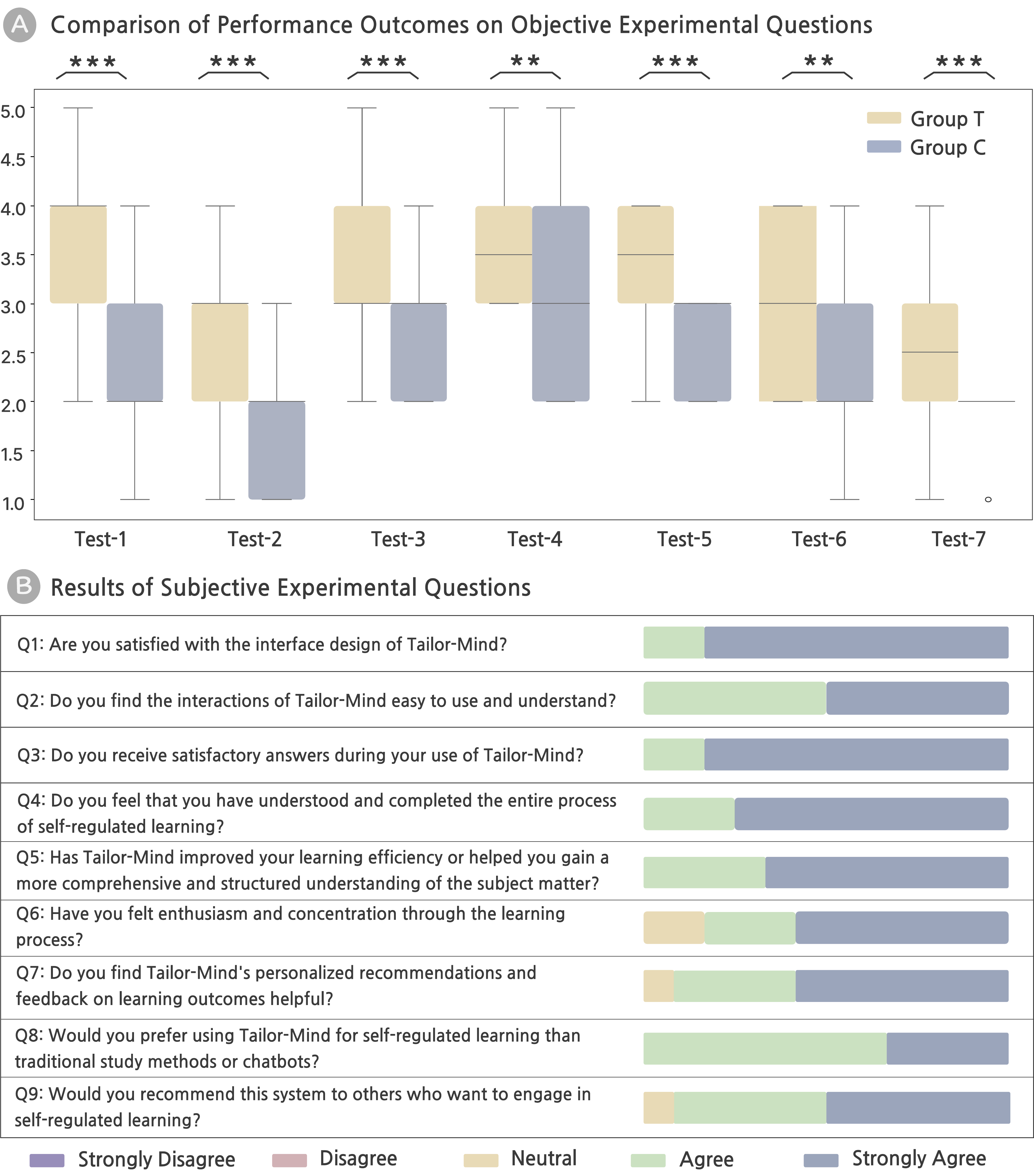}
  \caption{User study results. The box-plot (A) displays the performance outcomes of two groups on objective experimental questions. The number of asterisks (*) in the upper indicates the significance level of the test (*, **, *** for p <0.05, 0.01, and 0.005, respectively). (B) presents the detailed objective experimental questions and the corresponding distribution of satisfaction levels. Q1 through 3 pertain to Usability; Q4 to 6 focus on Effectiveness; Q7 indicates Customization; and Q8 and Q9 are about Recommendation. The results show that Tailor-Mind improves learning performance and receives good user feedback.}
  \label{fig: user study}
  \vspace{-0.5cm}
\end{figure}

\subsubsection{Results and Analysis}
\textbf{Enhanced efficiency in self-regulated learning.}
In addition to better performance on all objective questions, as shown in Fig.~\ref{fig: user study}A, we have observational results and a summary of interviews that corroborate this finding.
First, Tailor-Mind's responses are so precise and comprehensive that they minimize cognitive load while elaborating key concepts, making it particularly beneficial for novices. 
Participants T2, T7, and T10 expressed appreciation for examples provided in explanations that aided in understanding abstract concepts. 
T5 favored the inclusion of summaries with each answer, facilitating note-taking. 
T8 discussed Tailor-Mind in clarifying concepts such as layer normalization over batch normalization, \textit{"The answers were neither too detailed nor too vague for beginners."}
In contrast, participants in Group C posed more generalized questions, which resulted in information-laden answers and challenged beginners' comprehension. 
C3 and C6 noted that although responses were comprehensive, areas of confusion persisted, as ChatGPT seemed to assume they possessed prior knowledge.
\textcolor{revision}{C10 remained skeptical about the answers provided by ChatGPT and is constantly concerned that it may make mistakes.}
We also observed that C2 and C12 twice sought simpler explanations of previous responses in more accessible language.
Meanwhile, visualization also plays a crucial role in comprehension.
T8 initially found network structures complex but observed enhanced hierarchical understanding after switching to the ``dagre'' layout.
% T5 mentioned how Knowledge MindMap linked existing knowledge with new concepts, facilitating a more cohesive knowledge network and integrating previously fragmented information.
Second, questions level, one of the observational metrics, revealed that Tailor-Mind facilitated participants in delving into more profound questions.
T2 was impressed by the quality of the recommended questions, highlighting the model's ability to prompt multi-faceted questions and deeper exploration of concepts and their interrelations.
T4 commented that \textit{"I felt inspired by recommended questions to initiate spontaneous questions."}

\textbf{Improved learning habits and promotion of iterative learning.}
On one hand, Tailor-Mind guided participants toward a more structured learning habit. 
Group T was compelled to understand and establish a learning plan, in contrast to Group C, where only four participants spontaneously engaged in planning.
Interview results indicated that understanding key and difficult points aided in setting objectives. 
T4 and T6 found it easier to grasp key points from File Preview.
T2 appreciated the categorization and basic segmentation of knowledge points, \textit{"The system recommended a learning sequence that follows a logical progression, which could mitigate issues of missing prerequisite knowledge and was recognized only after I completed all the tasks."}.
Conversely, Group C largely depended on the learning material to question and determine their learning needs, and they were often unclear about what needed to be studied. 
Their questions were typically raised in response to confusion about GPT's prior answers, guiding subsequent inquiry.
On the other hand, the whole process encouraged participants in Group T to reflect not only in the self-reflection stage.
T2, T7, and T8 spontaneously conducted self-assessments, validating their knowledge through recommended complex questions before the self-reflection stage.
T6 appreciated the final test, which highlighted areas of misunderstanding despite their initial confidence in comprehension. 
\textit{"Many test questions are designed from a speculative perspective, presenting issues that may be classified as 'partially true'."}
\textcolor{revision}{The self-reflection phase also facilitated participants' planning for further learning. Both T10 and T12 developed clear plans for their upcoming learning goals.}
T7, a participant with good self-learning habits, used spontaneous questions to verify whether he understood the knowledge correctly, thus further consolidating or correcting his understanding of knowledge points.
\textit{"Through self-assessment and new learning paths, I found mistakes in questions about Layer Normalization. I need to consolidate its principles and functions further."} \textcolor{revision}{In contrast, five participants in Group C were involved in self-reflection.}

\textbf{Enhanced user engagement and facilitation of active learning.}
The exit questionnaire, depicted in Fig.~\ref{fig: user study}B, consisted of nine evaluation questions from four aspects. 
Tailor-Mind received high ratings in these four aspects. Participants expressed willingness to continue learning with Tailor-Mind and recommended it to other beginners.
At the same time, we found that the process of users formulating questions became simplified.
An automatic array of recommended questions rendered it more user-friendly and encouraged participants to engage more readily.
This not only strengthened their curiosity about exploring these questions but also propelled them to seek out and delve into novel knowledge areas independently.
T4 shared a sense of accomplishment upon encountering and accurately responding to a question about the differences between ``positional encoding'' and ``one-hot encoding'' in the self-reflection stage.
T7 acknowledged an initial reluctance towards adopting new methodologies. 
For him, determining where to start and sustaining motivation posed considerable challenges. 
However, Tailor-Mind made the learning journey more seamless and rewarding, significantly enhancing T7's enthusiasm for delving deeper into uncharted knowledge. 
This shift not only reflected T7's growing competence in navigating the system but also underscored the system's role in nurturing an enduring passion for learning.
\textcolor{revision}{Additionally, we observed that most participants in Group C complained about the slow response of GPT, and many did not know how to pose questions effectively.}

\section{Discussion} 
\label{sec:discussion}
% This section begins with a discussion from the perspective of Tailor-Mind and then extends it to our conceptual framework.
\subsection{Discussion on Tailor-Mind}
\textbf{Generalizability of Tailor-Mind.} 
Tailor-Mind can generalize to self-learning in all disciplines with structured knowledge. 
For any discipline and learner, maintaining refining knowledge structures is a general requirement of scientific self-learning, which is ensured by our visualization system and tuning methods. 
Currently, LLMs have uncertain capabilities in extracting semantic and structural information from images, audio, and video materials. 
Hence, Tailor-Mind does not support learning from these types of content \textcolor{revision}{and is not connected to web resources.}
However, with rapid advancements in multimodal large model capabilities, Tailor-Mind's framework and underlying fine-tuning mechanisms can be flexibly extended by converting non-text data into text format~\cite{yao2024effectiveness} \textcolor{revision}{and utilizing online data}.

\textbf{Enhancement for pipeline.}
The SRL pipeline's three stages offer potential for technical enhancements. 
In the forethought phase, future integration with web searches could complement question list recommendations, broadening the spectrum of learning resources. 
The performance phase could see the automatic optimization of human-recorded notes and their incorporation into a knowledge map. 
Meanwhile, the self-reflection stage might expand to include various self-assessment forms, such as error correction during learning. 
Additionally, employing multiple agents~\cite{park2023generative} throughout the whole SRL process could enrich the learning experience, with each agent assuming specific roles like tutoring, concentration monitoring, and incentives for learning.
The single-person SRL process can be further extended to community learning to obtain a wider range of learning experiences and effects.

\subsection{Discussion on General Framework}
\textbf{Generalizability of the framework.}  
Our frameworks involves specific domain, visualization system, user interactions, and LLMs.
% 我们的框架涉及到特定领域、可视化系统、用户交互以及LLM。
% domain features
% 适用：高度专业化（信息密集，知识具有广度和深度）；高结构性（具有定义明确的规则/标准以及领域专业知识）；问题解决与决策导向的，继而要求以用户为导向
% 不适用：高度主观的，开放性的领域；无统一明确的结构/规范；高情感依赖的
% 从适用性出发，我们的框架适合的领域是高度专业化的，信息密集且知识具有广度和深度；同时，领域具有高结构性的特征，有定义明确的规则与标准；领域任务为问题解决或决策为导向会更适用，继而要求这些问题的处理过程以用户为中心展开。
% 因此，我们框架的局限性在于处理尤其主观的、没有明确规范且具有较强情感依赖的领域任务，例如艺术创意、开放性实验研究等。
\textcolor{revision}{(1)~Specific domain: 
Our framework is well-suited for domains that are highly specialized, information-dense, and characterized by broad and deep knowledge. These domains are also highly structured, with well-defined rules and standards, and their tasks are problem-solving or decision-making oriented, requiring a user-centered approach. 
Therefore, the limitations of our framework lie in handling tasks that are highly subjective, lack clear norms, and have strong emotional dependencies, such as artistic creativity and open-ended experimental research.}
% visualization system
% interactive + VA
\textcolor{revision}{(2)~Visualization systems:
The well-defined functions of views in visualization systems ensure compatibility with our framework. In this work, we designed the visualization system to prioritize user interaction.} We believe that visual analytic (VA) systems focused on data analysis can also benefit from our framework. VA systems designed for text data can easily standardize requirements and domain knowledge into text-based fine-tuning tasks. For VA systems dealing with non-textual data, our framework could adapt as tuning methods evolve to teach models to analyze data, such as invoking numerical analysis APIs.
% LLM
\textcolor{revision}{(3)~LLMs:}
Although we verified the process on a relatively small open-source model, the entire process is also applicable to stronger closed-source models, such as GPT-4, since it can be fine-tuned. 
In specific applications of the framework, it is crucial to consider the trade-offs between model performance and access costs, ensuring the selection of the most suitable model for the given task.

\textbf{Iterative and customized fine-tuning.}
Considering the variations in users with different levels of knowledge and behavioral preferences, we find that personalized and customized fine-tuning could be beneficial. 
However, this process requires an initial automatic assessment of the user's knowledge level regarding the material, followed by adjusting the fine-tuning tasks based on that knowledge level, making automation challenging and costly. 
We believe that current fine-tuning based on the behavioral patterns of the majority of users can meet most user needs.
Although the information may be slightly overwhelming for beginners, interactive guidance in the interface can mitigate this issue.

% 毋庸置疑的是，微调的方式一定程度上从根源规范了模型的输出，根据微调指令数据做到了一定程度的行为规范。这是我们想要做到的所见即所得的效果，一定程度上让模型的输出变得可控和稳定，一致性问题在指令微调的作用下得到了缓和。
% 但是由于我们并不可能将所有可能性的数据都构造并用于微调，同时生成模型均带有输出的随机性，模型还是会输出看似合理但胡说八道的回答——幻觉问题难以避免。
% 这是由于微调还是基于已有的大语言模型，没有进行预训练式的全方面训练。然而为一个领域问题完全训练一个专属的大模型是不实际的（浪费资源的）。
% 为了减少错误回答，在未来的工作中，除了让LLM进行自我反思，进行错误报告机制也是必要的，让模型决策变得更加用户可信。这需要我们能够精准及时的捕捉到LLM所犯的错误，并给出推荐的有效解决方案。除了记录错误数据日志的方式之外，多agent协作监督也变得流行。
% 同时，模型对用户交互的对齐还可以做的更加全面。除了现有的对于人类偏好对齐的一些工作外，还应该让用户参与到微调数据构造的过程中，支持用户反馈模型错误并纠正错误回答。
\textcolor{revision}{\textbf{Performances of fine-tuned LLM.}
Fine-tuning standardizes model outputs through instruction data. It aligns LLM outputs with latent behavioral norms embedded in the data, resulting in more controllable, stable, and consistent outputs~\cite{chen2023maybe}.
However, as we cannot consider all possible data scenarios, and given the inherent randomness in generative models, the model still produces seemingly plausible but nonsensical responses, known as ``hallucinations''~\cite{lei2023hallucination}.
% This stems from fine-tuning being based on pre-existing LLMs rather than comprehensive pre-training. 
Fully training a specialized LLM for every domain is impractical and resource-intensive~\cite{Zhang2023BalancingSA}.
To reduce erroneous outputs, future work should include not only a self-reflection mechanism~\cite{Shinn2023ReflexionAA} but also an error reporting system to enhance trustworthiness. This requires accurately and promptly identifying LLM errors and providing effective solutions. In addition to logging erroneous data, multi-agent collaborative supervision is also a viable solution.
Additionally, alignment with user interactions can be further improved. Beyond current efforts in aligning with human preferences~\cite{ouyang2022training}, users should participate in constructing tuning data by supporting feedback and correction of model errors.}

\textbf{Effectiveness and comprehensiveness of evaluation.}
Our evaluation includes multiple effects and comprehensive methods, yet may still involve limitations. 
For the quantitative evaluation of a model, a common approach in the NLP community for fine-tuning tasks in specialized domains, where GPT's output is considered as ground truth, is comparing it with open-source models. 
We adopt this experimental setup and extend the comparison to GPT-3.5, demonstrating the rigor of our evaluation. 
In evaluating the system, we noted differences in users' knowledge levels but did not conduct grouped experiments to assess the impact of different user profiles on system effectiveness.
Additionally, we did not examine the differences in system usage before and after fine-tuning by a large pool of users, considering the cost associated with controlling all variables for comparative experiments.
However, comparing the fine-tuned model and system against GPT-4 (a strong baseline) showcases our system's superiority.
\section{Conclusion} \label{sec: conclusion}
To summarize, we proposed a framework integrating fine-tuned LLMs into visualization systems to achieve intelligent and interactive domain problem-solving. 
Based on this framework, we summarize a workflow for solving the three alignments among domain knowledge, visualization, interaction, and LLMs. 
To demonstrate the application of our framework, we introduce Tailor-Mind, an interactive intelligent visualization system. 
Following a detailed SRL pipeline, we designed fine-tuning data to improve intelligent decision-making and personalized recommendations. 
Through two usage scenarios, we illustrated that Tailor-Mind is suitable for beginners and aids in knowledge consolidation.
Model performance evaluations and user studies confirmed that Tailor-Mind is effective in promoting the scientific, active, and iterative SRL, which also validates the proposed framework and workflow.
\end{spacing}

\acknowledgments{%
The authors wish to thank Professor Yan Ding and her team from the Institute of Higher Education, Fudan University, for valuable feedback on this project.
This work is supported by Natural Science Foundation of China
(NSFC No.62202105, 62102323) and Shanghai Municipal Science and Technology Major Project (2021SHZDZX0103).
}

\bibliographystyle{./style/abbrv-doi-hyperref}
\normalem
\bibliography{main}

\begin{thebibliography}{10}

\bibitem{andrienko2018viewing}
N.~Andrienko, T.~Lammarsch, G.~Andrienko, G.~Fuchs, D.~Keim, S.~Miksch, and A.~Rind.
\newblock Viewing {Visual} {Analytics} as {Model} {Building}.
\newblock {\em Computer Graphics Forum}, 37(6):275--299, 2018. \href{https://doi.org/10.1111/cgf.13324}
{doi: {{%
10\hspace{.1pt}\discretionary{.}{%
}{.}\hspace{.4pt}1111\discretionary{/}{%
}{/}cgf\hspace{.1pt}\discretionary{.}{%
}{.}\hspace{.4pt}13324}}}


\bibitem{baladon2023retuyt}
A.~Balad{\'o}n, I.~Sastre, L.~Chiruzzo, and A.~Ros{\'a}.
\newblock {RETUYT}-{I}n{C}o at {BEA} 2023 shared task: Tuning open-source {LLM}s for generating teacher responses.
\newblock In E.~Kochmar, J.~Burstein, A.~Horbach, R.~Laarmann-Quante, N.~Madnani, A.~Tack, V.~Yaneva, Z.~Yuan, and T.~Zesch, eds., {\em Proceedings of the 18th Workshop on Innovative Use of NLP for Building Educational Applications (BEA 2023)}, pp. 756--765. Association for Computational Linguistics, Toronto, Canada, July 2023. \href{https://doi.org/10.18653/v1/2023.bea-1.61}
{doi: {{%
10\hspace{.1pt}\discretionary{.}{%
}{.}\hspace{.4pt}18653\discretionary{/}{%
}{/}v1\discretionary{/}{%
}{/}2023\hspace{.1pt}\discretionary{.}{%
}{.}\hspace{.4pt}bea\discretionary{%
}{-}{-}1\hspace{.1pt}\discretionary{.}{%
}{.}\hspace{.4pt}61}}}


\bibitem{chen2023maybe}
H.~Chen, Y.~Zhang, Q.~Zhang, H.~Yang, X.~Hu, X.~Ma, Y.~Yanggong, and J.~Zhao.
\newblock Maybe only 0.5\% data is needed: A preliminary exploration of low training data instruction tuning, 2023. \href{https://doi.org/10.48550/arXiv.2305.09246}
{doi: {{%
10\hspace{.1pt}\discretionary{.}{%
}{.}\hspace{.4pt}48550\discretionary{/}{%
}{/}arXiv\hspace{.1pt}\discretionary{.}{%
}{.}\hspace{.4pt}2305\hspace{.1pt}\discretionary{.}{%
}{.}\hspace{.4pt}09246}}}


\bibitem{chen2023program}
W.~Chen, X.~Ma, X.~Wang, and W.~W. Cohen.
\newblock Program of thoughts prompting: Disentangling computation from reasoning for numerical reasoning tasks, 2023. \href{https://doi.org/10.48550/arXiv.2211.12588}
{doi: {{%
10\hspace{.1pt}\discretionary{.}{%
}{.}\hspace{.4pt}48550\discretionary{/}{%
}{/}arXiv\hspace{.1pt}\discretionary{.}{%
}{.}\hspace{.4pt}2211\hspace{.1pt}\discretionary{.}{%
}{.}\hspace{.4pt}12588}}}


\bibitem{choi2022algosolve}
K.~Choi, H.~Shin, M.~Xia, and J.~Kim.
\newblock Algosolve: Supporting subgoal learning in algorithmic problem-solving with learnersourced microtasks.
\newblock In {\em Proceedings of the 2022 CHI Conference on Human Factors in Computing Systems}, pp. 1--16, 2022. \href{https://doi.org/10.1145/3491102.3501917}
{doi: {{%
10\hspace{.1pt}\discretionary{.}{%
}{.}\hspace{.4pt}1145\discretionary{/}{%
}{/}3491102\hspace{.1pt}\discretionary{.}{%
}{.}\hspace{.4pt}3501917}}}


\bibitem{crow2018intelligent}
T.~Crow, A.~Luxton-Reilly, and B.~Wuensche.
\newblock Intelligent tutoring systems for programming education: a systematic review.
\newblock In {\em Proceedings of the 20th Australasian Computing Education Conference}, pp. 53--62, 2018. \href{https://doi.org/10.1145/3160489.3160492}
{doi: {{%
10\hspace{.1pt}\discretionary{.}{%
}{.}\hspace{.4pt}1145\discretionary{/}{%
}{/}3160489\hspace{.1pt}\discretionary{.}{%
}{.}\hspace{.4pt}3160492}}}


\bibitem{dan2023educhat}
Y.~Dan, Z.~Lei, Y.~Gu, Y.~Li, J.~Yin, J.~Lin, L.~Ye, Z.~Tie, Y.~Zhou, Y.~Wang, A.~Zhou, Z.~Zhou, Q.~Chen, J.~Zhou, L.~He, and X.~Qiu.
\newblock Educhat: A large-scale language model-based chatbot system for intelligent education, 2023. \href{https://doi.org/10.48550/arXiv.2308.02773}
{doi: {{%
10\hspace{.1pt}\discretionary{.}{%
}{.}\hspace{.4pt}48550\discretionary{/}{%
}{/}arXiv\hspace{.1pt}\discretionary{.}{%
}{.}\hspace{.4pt}2308\hspace{.1pt}\discretionary{.}{%
}{.}\hspace{.4pt}02773}}}


\bibitem{du2023calla}
Y.~Du, S.~Zhao, Y.~Chen, R.~Bai, J.~Liu, H.~Wu, H.~Wang, and B.~Qin.
\newblock The calla dataset: Probing llms' interactive knowledge acquisition from chinese medical literature, 2023. \href{https://doi.org/10.48550/arXiv.2309.04198}
{doi: {{%
10\hspace{.1pt}\discretionary{.}{%
}{.}\hspace{.4pt}48550\discretionary{/}{%
}{/}arXiv\hspace{.1pt}\discretionary{.}{%
}{.}\hspace{.4pt}2309\hspace{.1pt}\discretionary{.}{%
}{.}\hspace{.4pt}04198}}}


\bibitem{fang2023ukp}
H.~Fang, H.~Puerto, and I.~Gurevych.
\newblock {UKP}-{SQ}u{ARE}: An interactive tool for teaching question answering.
\newblock In E.~Kochmar, J.~Burstein, A.~Horbach, R.~Laarmann-Quante, N.~Madnani, A.~Tack, V.~Yaneva, Z.~Yuan, and T.~Zesch, eds., {\em Proceedings of the 18th Workshop on Innovative Use of NLP for Building Educational Applications (BEA 2023)}, pp. 195--204. Association for Computational Linguistics, Toronto, Canada, July 2023. \href{https://doi.org/10.18653/v1/2023.bea-1.17}
{doi: {{%
10\hspace{.1pt}\discretionary{.}{%
}{.}\hspace{.4pt}18653\discretionary{/}{%
}{/}v1\discretionary{/}{%
}{/}2023\hspace{.1pt}\discretionary{.}{%
}{.}\hspace{.4pt}bea\discretionary{%
}{-}{-}1\hspace{.1pt}\discretionary{.}{%
}{.}\hspace{.4pt}17}}}


\bibitem{gao2023transforlearn}
L.~Gao, Z.~Shao, Z.~Luo, H.~Hu, C.~Turkay, and S.~Chen.
\newblock Transforlearn: Interactive visual tutorial for the transformer model.
\newblock {\em IEEE Transactions on Visualization and Computer Graphics}, 30(1):891--901, 2024. \href{https://doi.org/10.1109/TVCG.2023.3327353}
{doi: {{%
10\hspace{.1pt}\discretionary{.}{%
}{.}\hspace{.4pt}1109\discretionary{/}{%
}{/}TVCG\hspace{.1pt}\discretionary{.}{%
}{.}\hspace{.4pt}2023\hspace{.1pt}\discretionary{.}{%
}{.}\hspace{.4pt}3327353}}}


\bibitem{he2024large}
Q.~He, J.~Zeng, W.~Huang, L.~Chen, J.~Xiao, Q.~He, X.~Zhou, J.~Liang, and Y.~Xiao.
\newblock Can large language models understand real-world complex instructions?
\newblock {\em Proceedings of the AAAI Conference on Artificial Intelligence}, 38(16):18188--18196, Mar. 2024. \href{https://doi.org/10.1609/aaai.v38i16.29777}
{doi: {{%
10\hspace{.1pt}\discretionary{.}{%
}{.}\hspace{.4pt}1609\discretionary{/}{%
}{/}aaai\hspace{.1pt}\discretionary{.}{%
}{.}\hspace{.4pt}v38i16\hspace{.1pt}\discretionary{.}{%
}{.}\hspace{.4pt}29777}}}


\bibitem{hou2024c2ideas}
Y.~Hou, M.~Yang, H.~Cui, L.~Wang, J.~Xu, and W.~Zeng.
\newblock C2ideas: Supporting creative interior color design ideation with a large language model.
\newblock In {\em Proceedings of the CHI Conference on Human Factors in Computing Systems}, CHI '24. Association for Computing Machinery, New York, NY, USA, 2024. \href{https://doi.org/10.1145/3613904.3642224}
{doi: {{%
10\hspace{.1pt}\discretionary{.}{%
}{.}\hspace{.4pt}1145\discretionary{/}{%
}{/}3613904\hspace{.1pt}\discretionary{.}{%
}{.}\hspace{.4pt}3642224}}}


\bibitem{hu2022lora}
E.~J. Hu, yelong shen, P.~Wallis, Z.~Allen-Zhu, Y.~Li, S.~Wang, L.~Wang, and W.~Chen.
\newblock Lo{RA}: Low-rank adaptation of large language models.
\newblock In {\em International Conference on Learning Representations}, 2022. \href{https://doi.org/10.48550/arXiv.2106.09685}
{doi: {{%
10\hspace{.1pt}\discretionary{.}{%
}{.}\hspace{.4pt}48550\discretionary{/}{%
}{/}arXiv\hspace{.1pt}\discretionary{.}{%
}{.}\hspace{.4pt}2106\hspace{.1pt}\discretionary{.}{%
}{.}\hspace{.4pt}09685}}}


\bibitem{lei2023hallucination}
L.~Huang, W.~Yu, W.~Ma, W.~Zhong, Z.~Feng, H.~Wang, Q.~Chen, W.~Peng, X.~Feng, B.~Qin, and T.~Liu.
\newblock A survey on hallucination in large language models: Principles, taxonomy, challenges, and open questions.
\newblock {\em ArXiv}, abs/2311.05232, 2023. \href{https://doi.org/10.48550/arXiv.2311.05232}
{doi: {{%
10\hspace{.1pt}\discretionary{.}{%
}{.}\hspace{.4pt}48550\discretionary{/}{%
}{/}arXiv\hspace{.1pt}\discretionary{.}{%
}{.}\hspace{.4pt}2311\hspace{.1pt}\discretionary{.}{%
}{.}\hspace{.4pt}05232}}}


\bibitem{huang2024plantography}
R.~Huang, H.~Lin, C.~Chen, K.~Zhang, and W.~Zeng.
\newblock Plantography: Incorporating iterative design process into generative artificial intelligence for landscape rendering.
\newblock In {\em Proceedings of the CHI Conference on Human Factors in Computing Systems}, CHI '24. Association for Computing Machinery, New York, NY, USA, 2024. \href{https://doi.org/10.1145/3613904.3642824}
{doi: {{%
10\hspace{.1pt}\discretionary{.}{%
}{.}\hspace{.4pt}1145\discretionary{/}{%
}{/}3613904\hspace{.1pt}\discretionary{.}{%
}{.}\hspace{.4pt}3642824}}}


\bibitem{huang2023ceval}
Y.~Huang, Y.~Bai, Z.~Zhu, J.~Zhang, J.~Zhang, T.~Su, J.~Liu, C.~Lv, Y.~Zhang, J.~Lei, Y.~Fu, M.~Sun, and J.~He.
\newblock C-{Eval}: {A} {Multi}-{Level} {Multi}-{Discipline} {Chinese} {Evaluation} {Suite} for {Foundation} {Models}.
\newblock {\em Advances in Neural Information Processing Systems}, 36:62991--63010, Dec. 2023. \href{https://doi.org/10.48550/arXiv.2305.08322}
{doi: {{%
10\hspace{.1pt}\discretionary{.}{%
}{.}\hspace{.4pt}48550\discretionary{/}{%
}{/}arXiv\hspace{.1pt}\discretionary{.}{%
}{.}\hspace{.4pt}2305\hspace{.1pt}\discretionary{.}{%
}{.}\hspace{.4pt}08322}}}


\bibitem{hwang2020vision}
G.-J. Hwang, H.~Xie, B.~W. Wah, and D.~Ga{\v{s}}evi{\'c}.
\newblock Vision, challenges, roles and research issues of artificial intelligence in education, 2020. \href{https://doi.org/10.1016/j.caeai.2020.100001}
{doi: {{%
10\hspace{.1pt}\discretionary{.}{%
}{.}\hspace{.4pt}1016\discretionary{/}{%
}{/}j\hspace{.1pt}\discretionary{.}{%
}{.}\hspace{.4pt}caeai\hspace{.1pt}\discretionary{.}{%
}{.}\hspace{.4pt}2020\hspace{.1pt}\discretionary{.}{%
}{.}\hspace{.4pt}100001}}}


\bibitem{ji2023towards}
Z.~Ji, T.~Yu, Y.~Xu, N.~Lee, E.~Ishii, and P.~Fung.
\newblock Towards mitigating {LLM} hallucination via self reflection.
\newblock In H.~Bouamor, J.~Pino, and K.~Bali, eds., {\em Findings of the Association for Computational Linguistics: EMNLP 2023}, pp. 1827--1843. Association for Computational Linguistics, Singapore, 2023. \href{https://doi.org/10.18653/v1/2023.findings-emnlp.123}
{doi: {{%
10\hspace{.1pt}\discretionary{.}{%
}{.}\hspace{.4pt}18653\discretionary{/}{%
}{/}v1\discretionary{/}{%
}{/}2023\hspace{.1pt}\discretionary{.}{%
}{.}\hspace{.4pt}findings\discretionary{%
}{-}{-}emnlp\hspace{.1pt}\discretionary{.}{%
}{.}\hspace{.4pt}123}}}


\bibitem{yao2024effectiveness}
Y.~Jiang, X.~Yan, G.-P. Ji, K.~Fu, M.~Sun, H.~Xiong, D.-P. Fan, and F.~Khan.
\newblock Effectiveness assessment of recent large vision-language models.
\newblock {\em Visual Intelligence}, 2, 06 2024. \href{https://doi.org/10.1007/s44267-024-00050-1}
{doi: {{%
10\hspace{.1pt}\discretionary{.}{%
}{.}\hspace{.4pt}1007\discretionary{/}{%
}{/}s44267\discretionary{%
}{-}{-}024\discretionary{%
}{-}{-}00050\discretionary{%
}{-}{-}1}}}


\bibitem{kasneci2023chatgpt}
E.~Kasneci, K.~Se{\ss}ler, S.~K{\"u}chemann, M.~Bannert, D.~Dementieva, F.~Fischer, U.~Gasser, G.~Groh, S.~G{\"u}nnemann, E.~H{\"u}llermeier, et~al.
\newblock Chatgpt for good? on opportunities and challenges of large language models for education.
\newblock {\em Learning and individual differences}, 103:102274, 2023. \href{https://doi.org/10.1016/j.lindif.2023.102274}
{doi: {{%
10\hspace{.1pt}\discretionary{.}{%
}{.}\hspace{.4pt}1016\discretionary{/}{%
}{/}j\hspace{.1pt}\discretionary{.}{%
}{.}\hspace{.4pt}lindif\hspace{.1pt}\discretionary{.}{%
}{.}\hspace{.4pt}2023\hspace{.1pt}\discretionary{.}{%
}{.}\hspace{.4pt}102274}}}


\bibitem{Keim2008visual}
D.~Keim, G.~Andrienko, J.-D. Fekete, C.~Görg, J.~Kohlhammer, and G.~Melançon.
\newblock Visual {Analytics}: {Definition}, {Process}, and {Challenges}.
\newblock In A.~Kerren, J.~T. Stasko, J.-D. Fekete, and C.~North, eds., {\em Information {Visualization}}, vol. 4950, pp. 154--175. Springer Berlin Heidelberg, Berlin, Heidelberg, 2008. \href{https://doi.org/10.1007/978-3-540-70956-5_7}
{doi: {{%
10\hspace{.1pt}\discretionary{.}{%
}{.}\hspace{.4pt}1007\discretionary{/}{%
}{/}978\discretionary{%
}{-}{-}3\discretionary{%
}{-}{-}540\discretionary{%
}{-}{-}70956\discretionary{%
}{-}{-}5\_7}}}


\bibitem{david2002bloom}
D.~R. Krathwohl.
\newblock A revision of bloom's taxonomy: An overview.
\newblock {\em Theory Into Practice}, 41(4):212--218, 2002. \href{https://doi.org/10.1207/s15430421tip4104_2}
{doi: {{%
10\hspace{.1pt}\discretionary{.}{%
}{.}\hspace{.4pt}1207\discretionary{/}{%
}{/}s15430421tip4104\_2}}}


\bibitem{lewis2020rag}
P.~Lewis, E.~Perez, A.~Piktus, F.~Petroni, V.~Karpukhin, N.~Goyal, H.~K\"{u}ttler, M.~Lewis, W.-t. Yih, T.~Rockt\"{a}schel, S.~Riedel, and D.~Kiela.
\newblock Retrieval-augmented generation for knowledge-intensive nlp tasks.
\newblock In {\em Proceedings of the 34th International Conference on Neural Information Processing Systems}, NIPS '20. Curran Associates Inc., Red Hook, NY, USA, 2020. \href{https://doi.org/10.48550/arXiv.2005.11401}
{doi: {{%
10\hspace{.1pt}\discretionary{.}{%
}{.}\hspace{.4pt}48550\discretionary{/}{%
}{/}arXiv\hspace{.1pt}\discretionary{.}{%
}{.}\hspace{.4pt}2005\hspace{.1pt}\discretionary{.}{%
}{.}\hspace{.4pt}11401}}}


\bibitem{li2023quantity}
M.~Li, Y.~Zhang, Z.~Li, J.~Chen, L.~Chen, N.~Cheng, J.~Wang, T.~Zhou, and J.~Xiao.
\newblock From quantity to quality: Boosting llm performance with self-guided data selection for instruction tuning, 2023. \href{https://doi.org/10.48550/arXiv.2308.12032}
{doi: {{%
10\hspace{.1pt}\discretionary{.}{%
}{.}\hspace{.4pt}48550\discretionary{/}{%
}{/}arXiv\hspace{.1pt}\discretionary{.}{%
}{.}\hspace{.4pt}2308\hspace{.1pt}\discretionary{.}{%
}{.}\hspace{.4pt}12032}}}


\bibitem{ling2023domain}
C.~Ling and et~al.
\newblock Domain specialization as the key to make large language models disruptive: A comprehensive survey, 2023. \href{https://doi.org/10.48550/arXiv.2305.18703}
{doi: {{%
10\hspace{.1pt}\discretionary{.}{%
}{.}\hspace{.4pt}48550\discretionary{/}{%
}{/}arXiv\hspace{.1pt}\discretionary{.}{%
}{.}\hspace{.4pt}2305\hspace{.1pt}\discretionary{.}{%
}{.}\hspace{.4pt}18703}}}


\bibitem{liu2022semiconductors}
D.~Y. Liu, L.~M. Xu, X.~M. Lin, X.~Wei, W.~J. Yu, Y.~Wang, and Z.~M. Wei.
\newblock Machine learning for semiconductors.
\newblock {\em Chip}, 1(4):100033, 2022. \href{https://doi.org/10.1016/j.chip.2022.100033}
{doi: {{%
10\hspace{.1pt}\discretionary{.}{%
}{.}\hspace{.4pt}1016\discretionary{/}{%
}{/}j\hspace{.1pt}\discretionary{.}{%
}{.}\hspace{.4pt}chip\hspace{.1pt}\discretionary{.}{%
}{.}\hspace{.4pt}2022\hspace{.1pt}\discretionary{.}{%
}{.}\hspace{.4pt}100033}}}


\bibitem{liu2021supporting}
J.~Liu, T.~Dwyer, G.~Tack, S.~Gratzl, and K.~Marriott.
\newblock Supporting the problem-solving loop: Designing highly interactive optimisation systems.
\newblock {\em IEEE Transactions on Visualization and Computer Graphics}, 27(2):1764--1774, 2020. \href{https://doi.org/10.1109/TVCG.2020.3030364}
{doi: {{%
10\hspace{.1pt}\discretionary{.}{%
}{.}\hspace{.4pt}1109\discretionary{/}{%
}{/}TVCG\hspace{.1pt}\discretionary{.}{%
}{.}\hspace{.4pt}2020\hspace{.1pt}\discretionary{.}{%
}{.}\hspace{.4pt}3030364}}}


\bibitem{liu2024how}
Y.~Liu, S.~Chen, H.~Cheng, M.~Yu, X.~Ran, A.~Mo, Y.~Tang, and Y.~Huang.
\newblock How ai processing delays foster creativity: Exploring research question co-creation with an llm-based agent.
\newblock In {\em Proceedings of the CHI Conference on Human Factors in Computing Systems}, CHI '24. Association for Computing Machinery, New York, NY, USA, 2024. \href{https://doi.org/10.1145/3613904.3642698}
{doi: {{%
10\hspace{.1pt}\discretionary{.}{%
}{.}\hspace{.4pt}1145\discretionary{/}{%
}{/}3613904\hspace{.1pt}\discretionary{.}{%
}{.}\hspace{.4pt}3642698}}}


\bibitem{lu2024agentlens}
J.~Lu, B.~Pan, J.~Chen, Y.~Feng, J.~Hu, Y.~Peng, and W.~Chen.
\newblock Agentlens: Visual analysis for agent behaviors in llm-based autonomous systems.
\newblock {\em IEEE Transactions on Visualization and Computer Graphics}, pp. 1--17, 2024. \href{https://doi.org/10.1109/TVCG.2024.3394053}
{doi: {{%
10\hspace{.1pt}\discretionary{.}{%
}{.}\hspace{.4pt}1109\discretionary{/}{%
}{/}TVCG\hspace{.1pt}\discretionary{.}{%
}{.}\hspace{.4pt}2024\hspace{.1pt}\discretionary{.}{%
}{.}\hspace{.4pt}3394053}}}


\bibitem{ma2023hypocompass}
Q.~Ma, H.~Shen, K.~Koedinger, and S.~T. Wu.
\newblock How to teach programming in the ai era? using llms as a teachable agent for debugging, 2024. \href{https://doi.org/10.1007/978-3-031-64302-6_19}
{doi: {{%
10\hspace{.1pt}\discretionary{.}{%
}{.}\hspace{.4pt}1007\discretionary{/}{%
}{/}978\discretionary{%
}{-}{-}3\discretionary{%
}{-}{-}031\discretionary{%
}{-}{-}64302\discretionary{%
}{-}{-}6\_19}}}


\bibitem{openai2023chatgpt}
OpenAI.
\newblock Chatgpt: Optimizing language models for dialogue.
\newblock \url{https://www.openai.com/chatgpt}, 2023.

\bibitem{openai2024gpt4}
OpenAI.
\newblock Gpt-4 technical report.
\newblock \url{https://cdn.openai.com/papers/gpt-4.pdf}, 2024.

\bibitem{ouyang2022training}
L.~Ouyang and et~al.
\newblock Training language models to follow instructions with human feedback, 2022. \href{https://doi.org/10.48550/arXiv.2203.02155}
{doi: {{%
10\hspace{.1pt}\discretionary{.}{%
}{.}\hspace{.4pt}48550\discretionary{/}{%
}{/}arXiv\hspace{.1pt}\discretionary{.}{%
}{.}\hspace{.4pt}2203\hspace{.1pt}\discretionary{.}{%
}{.}\hspace{.4pt}02155}}}


\bibitem{panadero2017review}
E.~Panadero.
\newblock A review of self-regulated learning: Six models and four directions for research.
\newblock {\em Frontiers in psychology}, 8:422, 2017. \href{https://doi.org/10.3389/fpsyg.2017.00422}
{doi: {{%
10\hspace{.1pt}\discretionary{.}{%
}{.}\hspace{.4pt}3389\discretionary{/}{%
}{/}fpsyg\hspace{.1pt}\discretionary{.}{%
}{.}\hspace{.4pt}2017\hspace{.1pt}\discretionary{.}{%
}{.}\hspace{.4pt}00422}}}


\bibitem{park2023generative}
J.~S. Park, J.~O'Brien, C.~J. Cai, M.~R. Morris, P.~Liang, and M.~S. Bernstein.
\newblock Generative agents: Interactive simulacra of human behavior.
\newblock In {\em Proceedings of the 36th Annual ACM Symposium on User Interface Software and Technology}, UIST '23. Association for Computing Machinery, New York, NY, USA, 2023. \href{https://doi.org/10.1145/3586183.3606763}
{doi: {{%
10\hspace{.1pt}\discretionary{.}{%
}{.}\hspace{.4pt}1145\discretionary{/}{%
}{/}3586183\hspace{.1pt}\discretionary{.}{%
}{.}\hspace{.4pt}3606763}}}


\bibitem{peng2023check}
B.~Peng, M.~Galley, P.~He, H.~Cheng, Y.~Xie, Y.~Hu, Q.~Huang, L.~Liden, Z.~Yu, W.~Chen, and J.~Gao.
\newblock Check your facts and try again: Improving large language models with external knowledge and automated feedback, 2023. \href{https://doi.org/10.48550/arXiv.2302.12813}
{doi: {{%
10\hspace{.1pt}\discretionary{.}{%
}{.}\hspace{.4pt}48550\discretionary{/}{%
}{/}arXiv\hspace{.1pt}\discretionary{.}{%
}{.}\hspace{.4pt}2302\hspace{.1pt}\discretionary{.}{%
}{.}\hspace{.4pt}12813}}}


\bibitem{peng2023storyfier}
Z.~Peng, X.~Wang, Q.~Han, J.~Zhu, X.~Ma, and H.~Qu.
\newblock Storyfier: Exploring vocabulary learning support with text generation models.
\newblock In {\em Proceedings of the 36th Annual ACM Symposium on User Interface Software and Technology}, UIST '23. Association for Computing Machinery, New York, NY, USA, 2023. \href{https://doi.org/10.1145/3586183.3606786}
{doi: {{%
10\hspace{.1pt}\discretionary{.}{%
}{.}\hspace{.4pt}1145\discretionary{/}{%
}{/}3586183\hspace{.1pt}\discretionary{.}{%
}{.}\hspace{.4pt}3606786}}}


\bibitem{qiu2023docflow}
R.~Qiu, Y.~Tu, Y.-S. Wang, P.-Y. Yen, and H.-W. Shen.
\newblock Docflow: A visual analytics system for question-based document retrieval and categorization.
\newblock {\em IEEE Transactions on Visualization and Computer Graphics}, 30(2):1533--1548, 2024. \href{https://doi.org/10.1109/TVCG.2022.3219762}
{doi: {{%
10\hspace{.1pt}\discretionary{.}{%
}{.}\hspace{.4pt}1109\discretionary{/}{%
}{/}TVCG\hspace{.1pt}\discretionary{.}{%
}{.}\hspace{.4pt}2022\hspace{.1pt}\discretionary{.}{%
}{.}\hspace{.4pt}3219762}}}


\bibitem{schick2023toolformer}
T.~Schick, J.~Dwivedi-Yu, R.~Dessì, R.~Raileanu, M.~Lomeli, L.~Zettlemoyer, N.~Cancedda, and T.~Scialom.
\newblock Toolformer: Language models can teach themselves to use tools, 2023. \href{https://doi.org/10.48550/arXiv.2302.04761}
{doi: {{%
10\hspace{.1pt}\discretionary{.}{%
}{.}\hspace{.4pt}48550\discretionary{/}{%
}{/}arXiv\hspace{.1pt}\discretionary{.}{%
}{.}\hspace{.4pt}2302\hspace{.1pt}\discretionary{.}{%
}{.}\hspace{.4pt}04761}}}


\bibitem{schick2021generating}
T.~Schick and H.~Sch{\"u}tze.
\newblock Generating datasets with pretrained language models, Nov. 2021. \href{https://doi.org/10.18653/v1/2021.emnlp-main.555}
{doi: {{%
10\hspace{.1pt}\discretionary{.}{%
}{.}\hspace{.4pt}18653\discretionary{/}{%
}{/}v1\discretionary{/}{%
}{/}2021\hspace{.1pt}\discretionary{.}{%
}{.}\hspace{.4pt}emnlp\discretionary{%
}{-}{-}main\hspace{.1pt}\discretionary{.}{%
}{.}\hspace{.4pt}555}}}


\bibitem{schmucker2023ruffle}
R.~Schmucker, M.~Xia, A.~Azaria, and T.~Mitchell.
\newblock Ruffle\&riley: Towards the automated induction of conversational tutoring systems, 2023. \href{https://doi.org/10.48550/arXiv.2310.01420}
{doi: {{%
10\hspace{.1pt}\discretionary{.}{%
}{.}\hspace{.4pt}48550\discretionary{/}{%
}{/}arXiv\hspace{.1pt}\discretionary{.}{%
}{.}\hspace{.4pt}2310\hspace{.1pt}\discretionary{.}{%
}{.}\hspace{.4pt}01420}}}


\bibitem{shi2023reverse}
D.~Shi, W.~Cui, D.~Huang, H.~Zhang, and N.~Cao.
\newblock Reverse-engineering information presentations: recovering hierarchical grouping from layouts of visual elements.
\newblock {\em Visual Intelligence}, 1, 06 2023. \href{https://doi.org/10.1007/s44267-023-00010-1}
{doi: {{%
10\hspace{.1pt}\discretionary{.}{%
}{.}\hspace{.4pt}1007\discretionary{/}{%
}{/}s44267\discretionary{%
}{-}{-}023\discretionary{%
}{-}{-}00010\discretionary{%
}{-}{-}1}}}


\bibitem{Shinn2023ReflexionAA}
N.~Shinn, B.~Labash, and A.~Gopinath.
\newblock Reflexion: an autonomous agent with dynamic memory and self-reflection.
\newblock {\em ArXiv}, 2023. \href{https://doi.org/10.48550/arXiv.2303.11366}
{doi: {{%
10\hspace{.1pt}\discretionary{.}{%
}{.}\hspace{.4pt}48550\discretionary{/}{%
}{/}arXiv\hspace{.1pt}\discretionary{.}{%
}{.}\hspace{.4pt}2303\hspace{.1pt}\discretionary{.}{%
}{.}\hspace{.4pt}11366}}}


\bibitem{strobelt2022interactive}
H.~Strobelt, A.~Webson, V.~Sanh, B.~Hoover, J.~Beyer, H.~Pfister, and A.~M. Rush.
\newblock Interactive and visual prompt engineering for ad-hoc task adaptation with large language models.
\newblock {\em IEEE Transactions on Visualization and Computer Graphics}, 29(1):1146--1156, 2023. \href{https://doi.org/10.1109/TVCG.2022.3209479}
{doi: {{%
10\hspace{.1pt}\discretionary{.}{%
}{.}\hspace{.4pt}1109\discretionary{/}{%
}{/}TVCG\hspace{.1pt}\discretionary{.}{%
}{.}\hspace{.4pt}2022\hspace{.1pt}\discretionary{.}{%
}{.}\hspace{.4pt}3209479}}}


\bibitem{tack2023bea}
A.~Tack, E.~Kochmar, Z.~Yuan, S.~Bibauw, and C.~Piech.
\newblock The {BEA} 2023 shared task on generating {AI} teacher responses in educational dialogues.
\newblock In E.~Kochmar, J.~Burstein, A.~Horbach, R.~Laarmann-Quante, N.~Madnani, A.~Tack, V.~Yaneva, Z.~Yuan, and T.~Zesch, eds., {\em Proceedings of the 18th Workshop on Innovative Use of NLP for Building Educational Applications (BEA 2023)}, pp. 785--795. Association for Computational Linguistics, Toronto, Canada, July 2023. \href{https://doi.org/10.18653/v1/2023.bea-1.64}
{doi: {{%
10\hspace{.1pt}\discretionary{.}{%
}{.}\hspace{.4pt}18653\discretionary{/}{%
}{/}v1\discretionary{/}{%
}{/}2023\hspace{.1pt}\discretionary{.}{%
}{.}\hspace{.4pt}bea\discretionary{%
}{-}{-}1\hspace{.1pt}\discretionary{.}{%
}{.}\hspace{.4pt}64}}}


\bibitem{thirunavukarasu2023large}
A.~J. Thirunavukarasu, D.~S.~J. Ting, K.~Elangovan, L.~Gutierrez, T.~F. Tan, and D.~S.~W. Ting.
\newblock Large language models in medicine.
\newblock {\em Nature Medicine}, 29(8):1930--1940, Aug. 2023. \href{https://doi.org/10.1038/s41591-023-02448-8}
{doi: {{%
10\hspace{.1pt}\discretionary{.}{%
}{.}\hspace{.4pt}1038\discretionary{/}{%
}{/}s41591\discretionary{%
}{-}{-}023\discretionary{%
}{-}{-}02448\discretionary{%
}{-}{-}8}}}


\bibitem{touvron2023llama}
H.~Touvron, T.~Lavril, G.~Izacard, X.~Martinet, M.-A. Lachaux, T.~Lacroix, B.~Rozière, N.~Goyal, E.~Hambro, F.~Azhar, A.~Rodriguez, A.~Joulin, E.~Grave, and G.~Lample.
\newblock Llama: Open and efficient foundation language models, 2023. \href{https://doi.org/10.48550/arXiv.2302.13971}
{doi: {{%
10\hspace{.1pt}\discretionary{.}{%
}{.}\hspace{.4pt}48550\discretionary{/}{%
}{/}arXiv\hspace{.1pt}\discretionary{.}{%
}{.}\hspace{.4pt}2302\hspace{.1pt}\discretionary{.}{%
}{.}\hspace{.4pt}13971}}}


\bibitem{wang2023huatuo}
H.~Wang, C.~Liu, N.~Xi, Z.~Qiang, S.~Zhao, B.~Qin, and T.~Liu.
\newblock Huatuo: Tuning llama model with chinese medical knowledge, 2023. \href{https://doi.org/10.48550/arXiv.2304.06975}
{doi: {{%
10\hspace{.1pt}\discretionary{.}{%
}{.}\hspace{.4pt}48550\discretionary{/}{%
}{/}arXiv\hspace{.1pt}\discretionary{.}{%
}{.}\hspace{.4pt}2304\hspace{.1pt}\discretionary{.}{%
}{.}\hspace{.4pt}06975}}}


\bibitem{wang2023knowledgetuning}
H.~Wang, S.~Zhao, Z.~Qiang, Z.~Li, N.~Xi, Y.~Du, M.~Cai, H.~Guo, Y.~Chen, H.~Xu, B.~Qin, and T.~Liu.
\newblock Knowledge-tuning large language models with structured medical knowledge bases for reliable response generation in chinese, 2023. \href{https://doi.org/10.48550/arXiv.2309.04175}
{doi: {{%
10\hspace{.1pt}\discretionary{.}{%
}{.}\hspace{.4pt}48550\discretionary{/}{%
}{/}arXiv\hspace{.1pt}\discretionary{.}{%
}{.}\hspace{.4pt}2309\hspace{.1pt}\discretionary{.}{%
}{.}\hspace{.4pt}04175}}}


\bibitem{wang2023commonsensevis}
X.~Wang, R.~Huang, Z.~Jin, T.~Fang, and H.~Qu.
\newblock Commonsensevis: Visualizing and understanding commonsense reasoning capabilities of natural language models.
\newblock {\em IEEE Transactions on Visualization and Computer Graphics}, p. 1–11, 2023. \href{https://doi.org/10.1109/tvcg.2023.3327153}
{doi: {{%
10\hspace{.1pt}\discretionary{.}{%
}{.}\hspace{.4pt}1109\discretionary{/}{%
}{/}tvcg\hspace{.1pt}\discretionary{.}{%
}{.}\hspace{.4pt}2023\hspace{.1pt}\discretionary{.}{%
}{.}\hspace{.4pt}3327153}}}


\bibitem{wang2023selfinstruct}
Y.~Wang, Y.~Kordi, S.~Mishra, A.~Liu, N.~A. Smith, D.~Khashabi, and H.~Hajishirzi.
\newblock Self-instruct: Aligning language models with self-generated instructions, 2023. \href{https://doi.org/10.48550/arXiv.2212.10560}
{doi: {{%
10\hspace{.1pt}\discretionary{.}{%
}{.}\hspace{.4pt}48550\discretionary{/}{%
}{/}arXiv\hspace{.1pt}\discretionary{.}{%
}{.}\hspace{.4pt}2212\hspace{.1pt}\discretionary{.}{%
}{.}\hspace{.4pt}10560}}}


\bibitem{wang2024virtuwander}
Z.~Wang, L.-P. Yuan, L.~Wang, B.~Jiang, and W.~Zeng.
\newblock Virtuwander: Enhancing multi-modal interaction for virtual tour guidance through large language models.
\newblock In {\em Proceedings of the CHI Conference on Human Factors in Computing Systems}, CHI '24. Association for Computing Machinery, New York, NY, USA, 2024. \href{https://doi.org/10.1145/3613904.3642235}
{doi: {{%
10\hspace{.1pt}\discretionary{.}{%
}{.}\hspace{.4pt}1145\discretionary{/}{%
}{/}3613904\hspace{.1pt}\discretionary{.}{%
}{.}\hspace{.4pt}3642235}}}


\bibitem{wang2023multilevel}
Z.~Wang, Q.~Zhang, S.-W. HU, H.~Yu, X.~Jin, Z.~Gong, and H.~Chen.
\newblock Multi-level protein structure pre-training via prompt learning.
\newblock In {\em The Eleventh International Conference on Learning Representations}, 2023.

\bibitem{wei2022finetuned}
J.~Wei, M.~Bosma, V.~Y. Zhao, K.~Guu, A.~W. Yu, B.~Lester, N.~Du, A.~M. Dai, and Q.~V. Le.
\newblock Finetuned language models are zero-shot learners, 2022. \href{https://doi.org/10.48550/arXiv.2109.01652}
{doi: {{%
10\hspace{.1pt}\discretionary{.}{%
}{.}\hspace{.4pt}48550\discretionary{/}{%
}{/}arXiv\hspace{.1pt}\discretionary{.}{%
}{.}\hspace{.4pt}2109\hspace{.1pt}\discretionary{.}{%
}{.}\hspace{.4pt}01652}}}


\bibitem{Wei2022ChainOT}
J.~Wei, X.~Wang, D.~Schuurmans, M.~Bosma, B.~Ichter, F.~Xia, E.~H. Chi, Q.~V. Le, and D.~Zhou.
\newblock Chain-of-thought prompting elicits reasoning in large language models.
\newblock In {\em Proceedings of the 36th International Conference on Neural Information Processing Systems}, NIPS '22. Curran Associates Inc., Red Hook, NY, USA, 2024. \href{https://doi.org/10.48550/arXiv.2201.11903}
{doi: {{%
10\hspace{.1pt}\discretionary{.}{%
}{.}\hspace{.4pt}48550\discretionary{/}{%
}{/}arXiv\hspace{.1pt}\discretionary{.}{%
}{.}\hspace{.4pt}2201\hspace{.1pt}\discretionary{.}{%
}{.}\hspace{.4pt}11903}}}


\bibitem{wu2024socrates}
G.~Wu, S.~Guo, J.~Hoffswell, G.~Y.-Y. Chan, R.~A. Rossi, and E.~Koh.
\newblock Socrates: Data story generation via adaptive machine-guided elicitation of user feedback.
\newblock {\em IEEE Transactions on Visualization and Computer Graphics}, 30(1):131--141, 2024. \href{https://doi.org/10.1109/TVCG.2023.3327363}
{doi: {{%
10\hspace{.1pt}\discretionary{.}{%
}{.}\hspace{.4pt}1109\discretionary{/}{%
}{/}TVCG\hspace{.1pt}\discretionary{.}{%
}{.}\hspace{.4pt}2023\hspace{.1pt}\discretionary{.}{%
}{.}\hspace{.4pt}3327363}}}


\bibitem{xi2023rise}
Z.~Xi and et~al.
\newblock The rise and potential of large language model based agents: A survey, 2023. \href{https://doi.org/10.48550/arXiv.2309.07864}
{doi: {{%
10\hspace{.1pt}\discretionary{.}{%
}{.}\hspace{.4pt}48550\discretionary{/}{%
}{/}arXiv\hspace{.1pt}\discretionary{.}{%
}{.}\hspace{.4pt}2309\hspace{.1pt}\discretionary{.}{%
}{.}\hspace{.4pt}07864}}}


\bibitem{Yang2023Baichuan2O}
A.~Yang and et~al.
\newblock Baichuan 2: Open large-scale language models, 2023. \href{https://doi.org/10.48550/arXiv.2309.10305}
{doi: {{%
10\hspace{.1pt}\discretionary{.}{%
}{.}\hspace{.4pt}48550\discretionary{/}{%
}{/}arXiv\hspace{.1pt}\discretionary{.}{%
}{.}\hspace{.4pt}2309\hspace{.1pt}\discretionary{.}{%
}{.}\hspace{.4pt}10305}}}


\bibitem{Yang2023RefGPTDG}
D.~Yang, R.~Yuan, Y.~Fan, Y.~Yang, Z.~Wang, S.~Wang, and H.~Zhao.
\newblock {R}ef{GPT}: Dialogue generation of {GPT}, by {GPT}, and for {GPT}.
\newblock In H.~Bouamor, J.~Pino, and K.~Bali, eds., {\em Findings of the Association for Computational Linguistics: EMNLP 2023}, pp. 2511--2535. Association for Computational Linguistics, Singapore, 2023. \href{https://doi.org/10.18653/v1/2023.findings-emnlp.165}
{doi: {{%
10\hspace{.1pt}\discretionary{.}{%
}{.}\hspace{.4pt}18653\discretionary{/}{%
}{/}v1\discretionary{/}{%
}{/}2023\hspace{.1pt}\discretionary{.}{%
}{.}\hspace{.4pt}findings\discretionary{%
}{-}{-}emnlp\hspace{.1pt}\discretionary{.}{%
}{.}\hspace{.4pt}165}}}


\bibitem{yang2023fingpt}
H.~Yang, X.-Y. Liu, and C.~D. Wang.
\newblock Fingpt: Open-source financial large language models, 2023. \href{https://doi.org/10.48550/arXiv.2306.06031}
{doi: {{%
10\hspace{.1pt}\discretionary{.}{%
}{.}\hspace{.4pt}48550\discretionary{/}{%
}{/}arXiv\hspace{.1pt}\discretionary{.}{%
}{.}\hspace{.4pt}2306\hspace{.1pt}\discretionary{.}{%
}{.}\hspace{.4pt}06031}}}


\bibitem{jingsi2023taoli}
J.~Yu, J.~Zhu, Y.~Wang, Y.~Liu, H.~Chang, J.~Nie, C.~Kong, R.~Chong, XinLiu, J.~An, L.~Lu, M.~Fang, and L.~Zhu.
\newblock Taoli llama.
\newblock \url{https://github.com/blcuicall/taoli}, 2023.

\bibitem{yue2023disclawllm}
S.~Yue, W.~Chen, S.~Wang, B.~Li, C.~Shen, S.~Liu, Y.~Zhou, Y.~Xiao, S.~Yun, X.~Huang, and Z.~Wei.
\newblock Disc-lawllm: Fine-tuning large language models for intelligent legal services, 2023. \href{https://doi.org/10.48550/arXiv.2309.11325}
{doi: {{%
10\hspace{.1pt}\discretionary{.}{%
}{.}\hspace{.4pt}48550\discretionary{/}{%
}{/}arXiv\hspace{.1pt}\discretionary{.}{%
}{.}\hspace{.4pt}2309\hspace{.1pt}\discretionary{.}{%
}{.}\hspace{.4pt}11325}}}


\bibitem{yue2023mammoth}
X.~Yue, X.~Qu, G.~Zhang, Y.~Fu, W.~Huang, H.~Sun, Y.~Su, and W.~Chen.
\newblock Mammoth: Building math generalist models through hybrid instruction tuning, 2023. \href{https://doi.org/10.48550/arXiv.2309.05653}
{doi: {{%
10\hspace{.1pt}\discretionary{.}{%
}{.}\hspace{.4pt}48550\discretionary{/}{%
}{/}arXiv\hspace{.1pt}\discretionary{.}{%
}{.}\hspace{.4pt}2309\hspace{.1pt}\discretionary{.}{%
}{.}\hspace{.4pt}05653}}}


\bibitem{Yuheng2024LEVA}
Z.~Yuheng, Y.~Zhang, Y.~Zhang, X.~Zhao, J.~Wang, Z.~Shao, C.~Turkay, and S.~Chen.
\newblock Leva: Using large language models to enhance visual analytics.
\newblock {\em IEEE transactions on visualization and computer graphics}, PP, 03 2024. \href{https://doi.org/10.1109/TVCG.2024.3368060}
{doi: {{%
10\hspace{.1pt}\discretionary{.}{%
}{.}\hspace{.4pt}1109\discretionary{/}{%
}{/}TVCG\hspace{.1pt}\discretionary{.}{%
}{.}\hspace{.4pt}2024\hspace{.1pt}\discretionary{.}{%
}{.}\hspace{.4pt}3368060}}}


\bibitem{zeng2023glmb}
A.~Zeng and et~al.
\newblock {GLM}-130b: An open bilingual pre-trained model.
\newblock In {\em The Eleventh International Conference on Learning Representations}, 2023. \href{https://doi.org/10.48550/arXiv.2210.02414}
{doi: {{%
10\hspace{.1pt}\discretionary{.}{%
}{.}\hspace{.4pt}48550\discretionary{/}{%
}{/}arXiv\hspace{.1pt}\discretionary{.}{%
}{.}\hspace{.4pt}2210\hspace{.1pt}\discretionary{.}{%
}{.}\hspace{.4pt}02414}}}


\bibitem{zhang2022fine-tuning}
H.~Zhang, G.~Li, J.~Li, Z.~Zhang, Y.~ZHU, and Z.~Jin.
\newblock Fine-{Tuning} {Pre}-{Trained} {Language} {Models} {Effectively} by {Optimizing} {Subnetworks} {Adaptively}.
\newblock In {\em Advances in {Neural} {Information} {Processing} {Systems}}, vol.~35, pp. 21442--21454. Curran Associates, Inc., 2022. \href{https://doi.org/10.48550/arXiv.2211.01642}
{doi: {{%
10\hspace{.1pt}\discretionary{.}{%
}{.}\hspace{.4pt}48550\discretionary{/}{%
}{/}arXiv\hspace{.1pt}\discretionary{.}{%
}{.}\hspace{.4pt}2211\hspace{.1pt}\discretionary{.}{%
}{.}\hspace{.4pt}01642}}}


\bibitem{zhang2023instruction}
S.~Zhang, L.~Dong, X.~Li, S.~Zhang, X.~Sun, S.~Wang, J.~Li, R.~Hu, T.~Zhang, F.~Wu, and G.~Wang.
\newblock Instruction tuning for large language models: A survey, 2023. \href{https://doi.org/10.48550/arXiv.2308.10792}
{doi: {{%
10\hspace{.1pt}\discretionary{.}{%
}{.}\hspace{.4pt}48550\discretionary{/}{%
}{/}arXiv\hspace{.1pt}\discretionary{.}{%
}{.}\hspace{.4pt}2308\hspace{.1pt}\discretionary{.}{%
}{.}\hspace{.4pt}10792}}}


\bibitem{Zhang2023BalancingSA}
Z.~Zhang, C.~Zheng, D.~Tang, K.~Sun, Y.~Ma, Y.~Bu, X.~Zhou, and L.~Zhao.
\newblock Balancing specialized and general skills in llms: The impact of modern tuning and data strategy.
\newblock {\em ArXiv}, abs/2310.04945, 2023. \href{https://doi.org/10.48550/arXiv.2310.04945}
{doi: {{%
10\hspace{.1pt}\discretionary{.}{%
}{.}\hspace{.4pt}48550\discretionary{/}{%
}{/}arXiv\hspace{.1pt}\discretionary{.}{%
}{.}\hspace{.4pt}2310\hspace{.1pt}\discretionary{.}{%
}{.}\hspace{.4pt}04945}}}


\bibitem{zheng2022telling}
C.~Zheng, D.~Wang, A.~Y. Wang, and X.~Ma.
\newblock Telling stories from computational notebooks: Ai-assisted presentation slides creation for presenting data science work.
\newblock In {\em Proceedings of the 2022 CHI Conference on Human Factors in Computing Systems}, CHI '22. Association for Computing Machinery, New York, NY, USA, 2022. \href{https://doi.org/10.1145/3491102.3517615}
{doi: {{%
10\hspace{.1pt}\discretionary{.}{%
}{.}\hspace{.4pt}1145\discretionary{/}{%
}{/}3491102\hspace{.1pt}\discretionary{.}{%
}{.}\hspace{.4pt}3517615}}}


\bibitem{Zheng2023JudgingLW}
L.~Zheng, W.-L. Chiang, Y.~Sheng, S.~Zhuang, Z.~Wu, Y.~Zhuang, Z.~Lin, Z.~Li, D.~Li, E.~Xing, H.~Zhang, J.~E. Gonzalez, and I.~Stoica.
\newblock Judging llm-as-a-judge with mt-bench and chatbot arena.
\newblock In {\em Advances in Neural Information Processing Systems}, vol.~36, pp. 46595--46623. Curran Associates, Inc., 2023. \href{https://doi.org/10.48550/arXiv.2306.05685}
{doi: {{%
10\hspace{.1pt}\discretionary{.}{%
}{.}\hspace{.4pt}48550\discretionary{/}{%
}{/}arXiv\hspace{.1pt}\discretionary{.}{%
}{.}\hspace{.4pt}2306\hspace{.1pt}\discretionary{.}{%
}{.}\hspace{.4pt}05685}}}


\bibitem{zhong2023agieval}
W.~Zhong, R.~Cui, Y.~Guo, Y.~Liang, S.~Lu, Y.~Wang, A.~Saied, W.~Chen, and N.~Duan.
\newblock Agieval: A human-centric benchmark for evaluating foundation models, 2023. \href{https://doi.org/10.48550/arXiv.2304.06364}
{doi: {{%
10\hspace{.1pt}\discretionary{.}{%
}{.}\hspace{.4pt}48550\discretionary{/}{%
}{/}arXiv\hspace{.1pt}\discretionary{.}{%
}{.}\hspace{.4pt}2304\hspace{.1pt}\discretionary{.}{%
}{.}\hspace{.4pt}06364}}}


\bibitem{zimmerman2000srl}
B.~J. Zimmerman.
\newblock Chapter 2 - attaining self-regulation: A social cognitive perspective.
\newblock In M.~Boekaerts, P.~R. Pintrich, and M.~Zeidner, eds., {\em Handbook of Self-Regulation}, pp. 13--39. Academic Press, San Diego, 2000. \href{https://doi.org/10.1016/B978-012109890-2/50031-7}
{doi: {{%
10\hspace{.1pt}\discretionary{.}{%
}{.}\hspace{.4pt}1016\discretionary{/}{%
}{/}B978\discretionary{%
}{-}{-}012109890\discretionary{%
}{-}{-}2\discretionary{/}{%
}{/}50031\discretionary{%
}{-}{-}7}}}


\end{thebibliography}

\appendix % You can use the `hideappendix` class option to skip everything after \appendix

\end{document}